\documentclass[preprint,showpacs,preprintnumbers,amsmath,amssymb]{revtex4}

\usepackage{graphicx}
\usepackage{dcolumn}
\usepackage{bm}

\usepackage[english]{babel}

\usepackage[final]{changes}

\definechangesauthor[Nasser Mohieddin Abukhdeir]{NMA}{red}

\begin{document}

\title{Theory and computation of directional nematic phase ordering} 

\author{Ezequiel R. Soul\'e}
\email{ersoule@fi.mdp.edu.ar}
\affiliation{Institute of Materials Science and Technology (INTEMA), University of Mar del Plata and National Research Council (CONICET), J. B. Justo 4302, 7600 Mar del Plata, Argentina}

\author{Nasser Mohieddin Abukhdeir}%
 \email{nasser.abukhdeir@mcgill.ca}
\author{Alejandro D. Rey}
\email{alejandro.rey@mcgill.ca}
\affiliation{Department of Chemical Engineering, McGill University, Montreal, Quebec H3A 2B2}

\date{\today}
\begin{abstract}
 A computational study of morphological instabilities of a two-dimensional nematic front under directional growth was performed using a Landau-de Gennes type quadrupolar tensor order parameter model for the first-order isotropic/nematic transition of 5CB (pentyl-cyanobiphenyl).  A previously derived energy balance, taking anisotropy into account, was utilized to account for latent heat and an imposed morphological gradient in the time-dependent model.  Simulations were performed using an initially homeotropic isotropic/nematic interface.  Thermal instabilities in both the linear and non-linear regimes were observed and compared to past experimental and theoretical observations.  A sharp-interface model for the study of linear morphological instabilities, taking into account additional complexity resulting from liquid crystalline order, was derived.  Results from the sharp-interface model were compared to those from full two-dimensional simulation identifying the specific limitations of simplified sharp-interface models for this liquid crystal system.  In the non-linear regime, secondary instabilities were observed to result in the formation of defects, interfacial heterogeneities, and bulk texture dynamics.
\end{abstract}

\pacs{61.30.-v, 64.70.M-, 83.80.Xz, 83.10.-y}

\maketitle

\section{Introduction} \label{sec:intro}

\begin{figure}[htp]
\begin{center}
\includegraphics[width=3.5in]{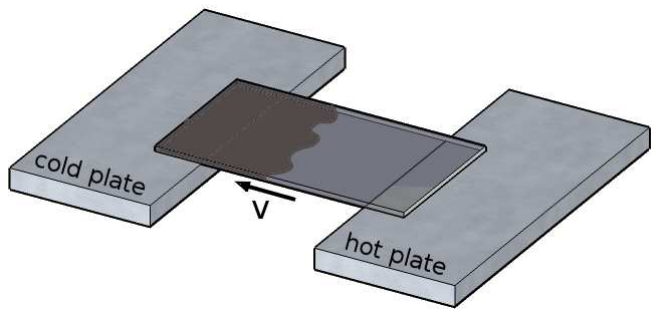} 
\end{center}
\caption{
Schematic of a directional growth experimental system where a slide/thin film enclosing a material is pulled at a constant velocity $v$ from a plate or oven at a temperature above the phase transition temperature to one that is below the phase transition temperature; the opaque portion of the slide represents the ordered/solid phase and the translucent portion represents the disordered/liquid. \label{fig:exp_dirsol}}
\end{figure}

Directional growth of materials undergoing a phase transition consists of pulling the material through a temperature gradient, usually from a higher to a lower temperature (relative to the phase transition temperature), such that an interface between the two phases is established in the central region of the sample.  A schematic of this setup is shown in Fig. \ref{fig:exp_dirsol}.  Directional growth configurations have practical uses such as producing single crystals with defined crystal orientation, producing eutectic composite materials, and as a purification method.  In addition to these applications, directional growth configurations are also widely used to study the fundamental nature of phase transitions.

One important characteristic present in general growth processes that involve phase transitions, and in the directional growth system in particular, is the occurrence of morphological instabilities \cite{Pelce2004}.  These instabilities arise in a great variety of materials undergoing different types of transitions, ranging from diffusive phase transitions to phase ordering.  The study of these instabilities has seen substantial advances since the seminal work of Mullins and Sekerka \cite{Mullins1963,Mullins1964,Coriell1985}.  Since then, some examples of systems that have been studied, in addition to the solidification of binary alloys \cite{Coriell1985,Wheeler1993,Braun1994,Wheeler1995,Sekerka2004}, include silicon wafer production in the semiconductor industry \cite{Sekerka2005}, polymer crystallization \cite{Xu2005a}, and liquid crystal growth \cite{Bechhoefer1989,Bechhoefer1995,Ignes-Mullol2000,Gomes2001}.

The fundamental understanding of morphological instabilities is important to the science of phase transformations.  These instabilities also affect the final material morphology (cellular, dendritic, textures in liquid crystals et al) and, subsequently, the structural and functional material properties.  For example, the internal structure of a liquid crystal spherulite controls the optical performance of polymer-dispersed liquid crystals.  Previous work \cite{Wincure2006,Wincure2007a} has shown that liquid crystal spherulite growth on the nanoscale leads to a series of orientational events, including defect nucleation and shape instabilities, showing that a better understanding of growth laws may be useful to develop improved optical materials. 

The study of directional growth in liquid crystalline material systems has been the focus of much previous work. These materials offer accessible time-scales for experimental observations as well as the combination of soft and anisotropic behavior.  The anisotropic character of liquid crystals results in many secondary instabilities, in addition to those described by Mullins and Sekerka \cite{Mullins1963,Mullins1964,Coriell1985}.  While liquid crystalline systems are templates for the overarching study of morphological instabilities, they are also pervasively used in industrial applications.  Their typical use in thin-film geometries, where large temperature gradients can be easily produced, places further importance on the need to characterize and understand the occurrence of morphological instabilities.

Much past experimental \cite{Bechhoefer1989,Ignes-Mullol2000,Gomes2001,Oswald1991,Simon1990,Mesquita1996} and theoretical \cite{Bechhoefer1995,Misbah1995} work has focused on studies of thermal and other types of instabilities of isotropic/nematic mesophase transitions in directional growth experiments.  This system is a convenient starting point since the nematic phase is the simplest of the liquid crystal mesophases.  Instabilities in smectic and columnar liquid crystals have been less studied \cite{Gonzalez-Cinca1998}.  Much of the work on morphological instabilities in nematic fronts has focused on explaining discrepancies between capillary lengths, determined experimentally \cite{Mesquita1996,Bechhoefer1989} and theoretically \cite{Bechhoefer1996,Gomes2001}.  Experimental investigations of the effects of convection and impurity concentration \cite{Bechhoefer1989} have been conducted, but the diverse set of possible morphological instabilities \cite{Oswald1987,Oswald1991,Bechhoefer1996,Ignes-Mullol2000} resulting from the inherent anisotropy and anchoring effects have not yet been completely explored.  The experimental study of these instabilities is inherently difficult due to the length and time scales involved.  On the other hand, computational studies are able to access these scales and shed light on the governing principles.

The overall mechanism which typically results in morphological instabilities is rooted in the existence of a temperature gradient, where the velocity of the interface $v$ is proportional to \cite{Wincure2006}: 
\begin{eqnarray} \label{eqn:velocity_interface}
v \propto \left|\Delta F \right| - C \nonumber\\
\left|\Delta F \right| \propto T_s-T
\end{eqnarray}
where $\left|\Delta F \right|$ is the free energy difference between the ordered/disordered phases, $C$ is the capillary force, $T_s$ is the coexistence temperature of the two phases, and $T$ is the temperature of the material.

\begin{figure}[htp]
\begin{center}
\includegraphics[width=3in]{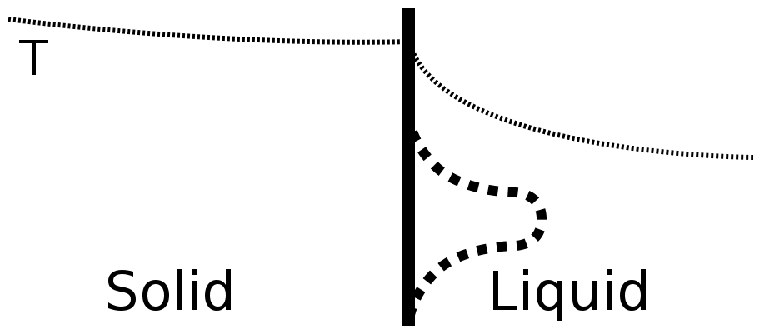} 
\end{center}
\caption{Schematic \added[NMA]{representation} of the Mullins-Sekerka instability \cite{Mullins1963,Mullins1964,Coriell1985} where the bold vertical line represents \added[NMA]{a flat} ordered/solid front growing into the disordered/liquid phase under an imposed temperature gradient\replaced[NMA]{, where the temperature profile is shown by the stippled line}{(which is superimposed on the schematic domain)}.  A perturbance in the front growth, the dotted line, is both inhibited by capillary forces and promoted by \replaced[NMA]{a greater decrease in free energy from phase transition as temperature decreases}{the decrease in free energy due to ordering/solidification} (see Eq. (\ref{eqn:velocity_interface})); note that the temperature of the ordered phase, in the vicinity of the interface (as shown), is greater than in the disordered phase due to latent heat effects. \label{fig:mulsek_inst}}
\end{figure}

When a shape perturbation arises in a moving interface, in the presence of a temperature gradient (refer to Fig. \ref{fig:mulsek_inst}), it finds a relatively increased $\left|\Delta F \right|$.  Subsequently, this promotes perturbation growth, increasing its relative velocity Eq. (\ref{eqn:velocity_interface}) (with respect to the unperturbed front).  In addition to this effect, the perturbation has curvature and thus a capillary force is exerted upon it, inhibiting growth.  This thermo-capillary competition results in the growth of perturbations with a wavelength (or radius of curvature) greater than a critical value, which decreases as the temperature gradient increases.

Although not studied in this work, it is also possible to observe instabilities induced by gradients in concentration.  This is referred to as constitutional super-cooling \cite{Flemmings1974}, and is the case in mixtures or when impurities are present. The ordered/solid phase has a lower equilibrium concentration of the solute or impurities than the disordered/liquid phase, thus the impurities are rejected to the disordered/liquid phase.  This results in the concentration of the impurity at the interface being higher than in the bulk disordered/liquid phase.  Any perturbation in the interface will find a lower concentration of impurity, resulting in a higher coexistence temperature and increased front velocity (refer to Eq. (\ref{eqn:velocity_interface})).  For a more in-depth explanation of morphological instabilities see Refs. \cite{Flemmings1974,Pimpinelli1998,Narayan2002,Sethna2007}.

\begin{figure}[htp]
\begin{center}
\includegraphics[width=3.5in]{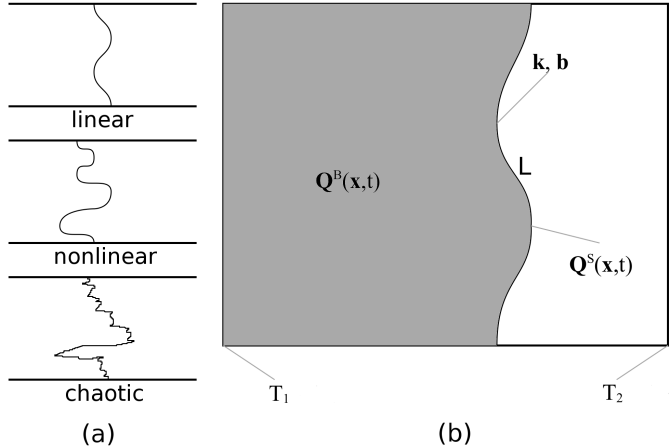} 
\end{center}
\caption{
a) schematics of the different regimes of morphological instabilities where the control variables include the pulling velocity $v$ and the imposed global temperature gradient $\Delta T / l$, where $l$ is the distance between the heat sources b) schematic of the liquid crystal physics taken into account in the non-isothermal tensorial Landau-de Gennes model \cite{Abukhdeir2008b} for the isotropic/nematic transition employed in this work: non-isothermal conditions ($T_1/T_2$ are the imposed boundary temperatures), bulk texture ($\bm{Q}^B$ is the tensor order parameter field in the bulk), interfacial gradients and heterogeneities ($\bm{Q}^S$ is the tensor order parameter field at the interface), heat of transition ($L$), and anisotropy/capillary forces ($\bm{k}/\bm{b}$ is the interface normal/curvature tensor). \label{fig:physics}}
\end{figure}

In a directional growth system, an appropriate temperature gradient, in the vicinity of the interface (\added[NMA]{see} Fig. \ref{fig:mulsek_inst}), can be produced via latent heat resulting from the phase transition.  Thus, even though the externally imposed temperature gradient \replaced[NMA]{stabilizes the interface}{is positive} (higher temperature in the disordered/liquid phase than in the ordered/solid phase), a \replaced[NMA]{destabilizing}{negative} temperature gradient can exist locally in the vicinity of the interface (depending on the magnitude of the heat of \added[NMA]{phase} transition). For some value of the \replaced[NMA]{externally}{external} imposed temperature gradient, the local gradient at the interface will be exactly zero, thus for external gradients greater than this value, the shape instability will not be observed. If the external gradient is lower than this value (or if it is negative), there will be some wavelength range for which the perturbations grow, so a shape instability is observed. At relatively low pulling velocity, stationary sinusoidal shape patterns are observed, produced by this temperature gradient. As velocity is increased, the patterns remain periodic but lose sinusoidal shape. At further increased velocities, nonlinear effects, including non-periodic/chaotic instabilities can be observed in some cases \cite{Bechhoefer1996} (see Fig. \ref{fig:physics}a). Finally, for very high velocities, a restabilization is possible which reforms a flat interface. 

Mullins and Sekerka were the first to model morphological instability using a sharp-interface model, in the linear regime, considering that growth was limited by diffusion \cite{Mullins1963,Mullins1964}.  In these models, the equations for heat or mass diffusion are solved in each phase, and the boundary is discontinuous.  Utilizing this formulation, dispersion diagrams, where the growth velocity versus wavelength are plotted, are able to be obtained analytically for a sinusoidal perturbation. This type of model has been extended to account for some types of nonlinear phenomena \cite{Bechhoefer1996,Xu1997,Misbah1995} and other effects, but are generally not feasible for use to model complex scenarios where nonlinear instabilities occur.

Phase field models \cite{Pismen2006} have also been used to model morphological instabilities.  These models inherently capture a more complete set of physics involved in that the material is modeled as a continuum.  Thus the interface is no longer assumed discontinuous and, in the case of liquid crystals, texturing processes can be resolved.  Furthermore, interfacial and bulk properties are determined implicitly by the model and its parameters.  A general comparison of both types of models (for a scalar phase field model) can be found in Ref. \cite{Braun1994}.

The Landau-de Gennes tensorial model for the isotropic/nematic transition \cite{deGennes1995} is one of the most effective theoretical approaches to capture the kinetics and dynamics of this transition at mesoscopic scales \cite{Gramsbergen1986,Singh2000}.  The phenomenological nature of this model is conducive to experimental validation and, as a result of these comparisons, this model has been shown to be relatively successful at capturing the physics of the isotropic/nematic transition \cite{Gramsbergen1986}.  The more coarse-grained approach of the Landau-type models, allows access to multi-scale phenomena which are relevant to experimental observations, but at a resolution unattainable other than through numerical simulation.  Thus, this model has been applied to study a broad range of phase-ordering phenomena, from use as a template to study the formation of the early universe \cite{Kibble2007} to the structure of liquid crystalline fibers, membranes, films, and drops \cite{Rey2007}.

With respect to this model of isotropic/nematic liquid crystalline transition, Fig. \ref{fig:physics}b elucidates the key physics captured in a schematic of  morphological instabilities in directional growth of nematic phase ordering; $\bm{Q}$ is the quadrupolar tensor order parameter, $\bm{k}$ is the outward unit normal, and $\bm{b}$ is the curvature tensor of the interface.  In nematic liquid crystals, the interface shape is coupled to the temperature and the order parameter $\bm{Q}$.  The coupling between the degree of phase ordering and temperature is implicitly accounted for in the Landau-de Gennes tensorial model.  The incorporation of a previously derived energy balance \cite{Abukhdeir2008b} accounts for the heat of transition and anisotropy in heat conduction (arising from the imposed boundary temperatures and the heat of phase ordering).  The $\bm{Q}$-tensor model implicitly incorporates the non-trivial couplings between growth, shape, and texturing dynamics.

The majority of theoretical approaches to the study of morphological instabilities in directional growth, both in the isotropic/nematic mesophase transition and the general case, utilize simple sharp-interface \cite{Bechhoefer1996} models and focus mainly on the linear regime.  The main advantage to this approach is that simple analytical solutions can be obtained and there is good agreement between the predictions of these types of models and experimental observations in the linear regime.  These models are not suitable for the study of instabilities in the non-linear regime, where much of the physics that is neglected in linearized sharp-interface models plays a role.  An example of these complex physics are, in the case of liquid crystals, defect formation and texturing processes, which have been experimentally observed \cite{Oswald1987,Oswald1991,Bechhoefer1996,Ignes-Mullol2000}. Thus, past theoretical work was focused on modeling the linear regime and some simple non-linear effects, but the complex texturing and defect formation in directional growth of liquid crystals has not been modelled before.  The state of the art in the study of this seminal problems is summarized in Chapter B.VI of ref. \cite{Oswald2005} where it is explained in detail that current simulation and theoretical work is limited to low-dimensional, director-type macroscopic models.  In the present case multi-scaling and multi-dimensionality are extended to the current computational limits.

The main objectives of this work on modeling the directional solidification of calamitic low molar mass nematic liquid crystals are:
\begin{itemize}
\item compare the $\mathbf{Q}(x,y,t)$ and $T(x,y,t)$ predictions of two-dimensional ($x,y$) transient simulation of the non-isothermal Landau-de Gennes model \cite{deGennes1995,Wincure2006,Abukhdeir2008b} results in the linear (small amplitude) regime with those from the standard analytical (sharp-interface) method. 
\item use the non-isothermal Landau-de Gennes model to simulate $\mathbf{Q}(x,y,t)$ and $T(x,y,t)$ in the nonlinear regime, where secondary instabilities result in texturing and defect formation. 
\end{itemize}
The main assumptions of this work are:
\begin{itemize}
\item the physical set-up and geometry correspond to the classical directional solidification experiment, as shown in Fig. \ref{fig:exp_dirsol}.
\item the liquid crystal is a pure thermotropic calamitic low molar mass material (see Table \ref{tab:parameters}).
\item directional growth is considered to be driven by an externally imposed temperature gradient.
\item thermal fluctuations are neglected.
\item constant physical properties corresponding to the 5CB (pentyl-cyanobiphenyl) liquid crystal \cite{Wincure2007a}, including thermal conductivity \cite{Pestov2003}, are used.
\item a periodic perturbation to the interface is assumed in the initial conditions.
\end{itemize} 
A full tensorial Landau-de Gennes model is used \cite{deGennes1995,Wincure2006} and with an applicable previously derived thermal energy balance \cite{Abukhdeir2008b}.  The nonlinear sharp-interface theory is not used because analytical solutions are infeasible when taking into account anisotropy, defect formation, and textures. 

\replaced[NMA]{This work is organized in three distinct parts introducing the relevant physics of nematic directional growth, modeling and simulation using the tensorial Landau-de Gennes model, and presentation of the results.  Sec. \ref{subsec_lcorder} presents a description of the isotropic/nematic phase transition and quadrupolar tensor order parameter used in the Landau-de Gennes model.  Sec. \ref{subsec:ord_model} presents a brief description of the Landau-de Gennes model and the nematic thermal energy balance (the latter accounts for dissipation, anisotropic heat conduction, and latent heat of transition).  Sec. \ref{subsec:simcond} describes the simulation conditions and complexities resulting in the use of the high-order Landau-de Gennes model.  Sec. \ref{sec:resdisc} is divided into two subsections presenting results of both linear and nonlinear regime morphological instabilities.  The linear regime results were found through analytical solutions of a derived sharp-interface model and through simulation of the higher-order Landau de Gennes model.  Nonlinear regime results were only accessible through simulation.  Finally, Sec. \ref{sec:conc} summarizes the findings and presents conclusions.  These conclusions focus on the limitations to sharp-interface model approaches compared to the Landau-de Gennes model.  Additionally, interesting results in the nonlinear regime are presented where a an interfacial reorientation is mechanism is found to occur through defect shedding.}{This work is organized as follows: Sec. \ref{subsec_lcorder} presents a description of the isotropic/nematic phase transition and quadrupolar tensor order parameter used in the Landau-de Gennes model, Sec. \ref{subsec:ord_model} presents a brief description of the Landau-de Gennes model and the nematic thermal energy balance (the latter accounts for dissipation, anisotropic heat conduction, and latent heat of transition), and Sec. \ref{sec:resdisc} presents the results that include linear and nonlinear regimes.}

\section{Background and theory}
\subsection{Liquid crystalline order} \label{subsec_lcorder}

Liquid crystalline phases or mesophases are materials which exhibit partial orientational and/or translational order.  They are composed of anisotropic molecules  which can be disc-like (discotic) or rod-like (calamitic) in shape.  Thermotropic liquid crystals are compounds that exhibit mesophase ordering in response to temperature changes.  Lyotropic liquid crystals that most greatly exhibit mesophase behavior in response to concentration changes.  Effects of pressure and external fields also influence mesophase behavior.  This work focuses the study of calamitic thermotropic liquid crystals which exhibit a first-order mesophase transition.

\begin{figure}[htp]
\begin{center}
\includegraphics[width=3.5in]{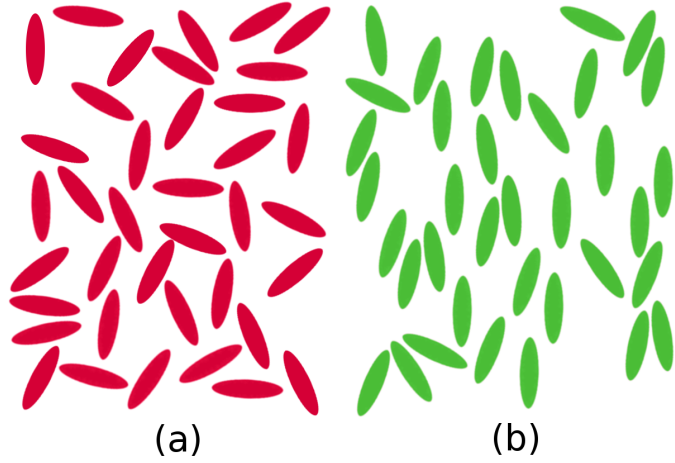} 
\end{center}
\caption{\added[NMA]{(Color online)} schematics of the a) isotropic and b) uniaxial nematic phases with rod-like mesogens. \label{figlcorder}}
\end{figure}

An unordered liquid, where there is neither orientational nor translational order (apart from an average intermolecular separation distance) of the molecules, is referred to as isotropic.  Uniaxial nematic liquid crystalline order in rod-like mesogens involves partial orientational order and positional disorder, where an average orientational axis, known as the director, is observed.  Schematic representations of these uniaxial nematic and isotropic ordering are shown in Fig. \ref{figlcorder}.

\subsection{Order parameters and the phenomenological model} \label{subsec:ord_model}

Theoretical characterization of nematic order is accomplished using an order parameter that adequately captures the physics involved.  This order parameter has an amplitude and phase associated with it.  In order to characterize the partial orientational order of the nematic phase, a second order symmetric traceless tensor can be used \cite{deGennes1995}:
\begin{equation} \label{eqnem_order_param}
\bm{Q} = S \left(\bm{nn} - \frac{1}{3} \bm{I}\right) + \frac{1}{3} P \left( \bm{mm} - \bm{ll}\right)
\end{equation}
where $\mathbf{n}/\mathbf{m}/\mathbf{l}$ are the eigenvectors of $\bm{Q}$, which characterize the average molecular orientational axes, and $S/P$ are scalars which represent the extent to which the molecules conform to the average orientational axes \cite{Rey2002,Yan2002,Rey2007}.  Uniaxial order is characterized by $S$ and $\bm{n}$, which correspond to the maximum eigenvalue (and its corresponding eigenvector) of $\bm{Q}$, $S= \frac{3}{2} \mu_n$.  Biaxial order is characterized by $P$ and $\bm{m}/\bm{l}$, which correspond to the lesser eigenvalues and eigenvectors, $P = -\frac{3}{2}\left(\mu_m - \mu_l\right)$.

The Landau-de Gennes model for the first order isotropic/nematic phase transition \cite{deGennes1995}:
\begin{eqnarray} \label{eq:free_energy_heterogeneous}
f - f_0 &=&\frac{1}{2} a \left(\bm{Q} : \bm{Q}\right) - \frac{1}{3} b \left(\bm{Q}\cdot\bm{Q}\right) : \bm{Q} + \frac{1}{4} c \left(\bm{Q} : \bm{Q}\right)^2 \nonumber\\
&& + \frac{1}{2} l_1 (\bm{\nabla} \bm{Q} \vdots \bm{\nabla} \bm{Q} ) + \frac{1}{2} l_2 \left( \nabla \cdot \bm{Q} \right) \cdot \left( \nabla \cdot \bm{Q} \right)\nonumber\\
&&+ \frac{1}{2} l_3 \bm{Q}:\left( \nabla \bm{Q} : \nabla \bm{Q} \right)
\end{eqnarray}

\begin{eqnarray} \label{eq:free_energy_heterogenous_coeffs}
a & = & a_0 (T - T_{NI}) \\
\end{eqnarray}
where $a$, $b$, $c$ are bulk parameters, $T_{NI}$ is the lower stability limit of the isotropic phase, and $l_1$, $l_2$, $l_3$ are the elastics constants. An equi-bend/splay assumption is used in this work, resulting in $l_1$, $l_2>0$ and $l_3=0$. This assumption is made based upon previous studies of nematic spherulite morphology resulting from the interplay of splay, twist, and bend elastic constants \cite{Wincure2006,Wincure2007,Wincure2007a,Wincure2007b}. Interfacial contributions implicitly result from the inclusion of the gradients terms in Eq. (\ref{eq:free_energy_heterogeneous}).  Detailed past work has studied the interfacial contributions of these gradient terms including anchoring, curvature, and heterogeneous effects. See to Refs. \cite{Wincure2006,Wincure2007,Wincure2007a,Wincure2007b} for a comprehensive study of these effects.

  The Landau-Ginzburg time dependent formulation \cite{Barbero2000} is used to minimize the free energy functional Eq. (\ref{eq:time_dep_formulation}) of the simulation volume:
\begin{eqnarray} \label{eq:time_dep_formulation}
F&=&\int_V f dV\\
\mu\frac{\partial \bm{Q}}{\partial t} &=& -\left[\frac{\delta F}{\delta \bm{Q}}\right]^{ST}
\end{eqnarray}
where $F$ is the total free energy, $\mu$ is the rotational viscosity and only the symmetric-traceless component of the functional derivative is utilized (denoted by the superscript \textit{ST}).
  The general differential energy balance, neglecting convection, is \cite{Abukhdeir2008b}:
\begin{equation} \label{eq:energy_balance}
\rho C_p \frac{\partial T}{\partial t} = \mu \frac{\partial \bm{Q}}{\partial t}:\frac{\partial\bm{Q}}{\partial t} + T \left\{ \frac{\partial}{\partial \bm{Q}} \frac{\partial f}{\partial T}\right\}^{ST}:\frac{\partial \bm{Q}}{\partial t} - \nabla \cdot \bm{q} 
\end{equation}
where $C_p$ is the specific heat and $\bm{q}$ is the total heat flux. The first right hand side term in Eq. \ref{eq:energy_balance} is dissipation \added[NMA]{due to nematic order dynamics}, the second \replaced[NMA]{the heat of transition to nematic order}{latent heat}, and the last thermal diffusion.  Temperature fluctuations are neglected in this model but could be incorporated via stochastic terms.  

The heat flux can be calculated from the anisotropic Fourier's law:
\begin{equation} \label{eq:fourier}
\bm{q} = −\bm{K}\cdot \nabla T   
\end{equation}                                                     
where the thermal conductivity tensor $\bm{K}$ is used due to the anisotropy of the nematic phase. The thermal conductivity tensor can be written as the sum of isotropic and anisotropic contributions:
\begin{equation} \label{eq:thermal_conductivity}
\bm{K} = k_{iso} \bm{\delta} + k_{an}\bm{Q} = \left(\frac{k_{\parallel}+2 k_{\perp}}{3}\right) \bm{\delta}+ \left(k_{\parallel}- k_{\perp}\right)\bm{Q}
\end{equation}   
where $k_{iso}$ and $k_{an}$ are the isotropic and anisotropic contributions to the thermal conductivities, and $k_{\parallel}$ and $k_{\perp}$ are the conductivities in the directions parallel and perpendicular to the director, respectively.

\subsection{Simulation method} \label{subsec:simcond}

\begin{figure}[htp]
\begin{center}
\includegraphics[width=2.5in]{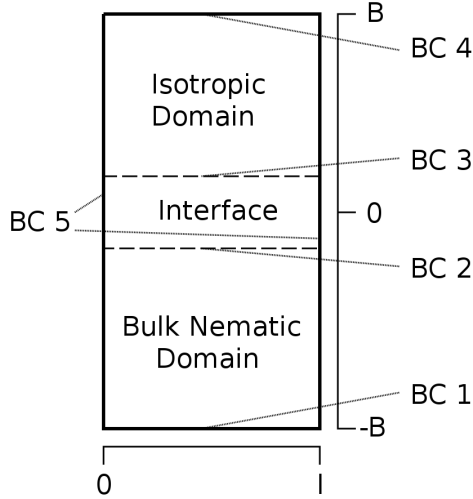} 
\end{center}
\caption{Schematic of the computational domain where the left sub-domain is the bulk homogeneously oriented bulk nematic domain, the center sub-domain encompasses the isotropic/nematic interface, the right sub-domain is fully isotropic, and the width/height of the domain are \added[NMA]{$2B$ and $l$}, respectively; numbering corresponds to the boundary conditions used where: 1) $T=T_1$ and Neumann conditions are used for $\bm{Q}$, 2) $T/\bm{Q}$ are coupled 3) $T$ is coupled and $\bm{Q}=0$, 4) $T=T_2$ and Neumann conditions are used for $\bm{Q}$, and 5) symmetry boundary conditions for $T/\bm{Q}$, see Eq. (\ref{eqn:symmetry_bcs}); note that in this work both $T_1$ and $T_2$ are below the bulk isotropic/nematic transition temperature and $T_2 < T_1$ due to computational limitations (see Sec. \ref{subsec:simcond}).\label{fig:compdomain}}
\end{figure}

A schematic of the geometry of the two-dimensional simulation domain and boundary condition types are shown in Fig. \ref{fig:compdomain}.  A central sub-domain, with a very refined mesh, is used in order to resolve the details at the nematic/isotropic interface.  Two outer sub-domains, with coarser meshes, are used to resolve the gradients in the order parameter in the bulk nematic phase and in temperature.  These bulk gradients are at length scales orders of magnitude greater than those at the interface.  Thus, the use of the fine mesh in these sub-domains is not necessary to resolve the textures. In the right sub-domain, a bulk isotropic phase is assumed (verified a posteriori), only the thermal energy balance Eq. (\ref{eq:energy_balance}) was solved.

\replaced[NMA]{Due to computational limitations the length scales accessible via simulation in the present work are on the order of microns.  Experimental observations of morphological instabilities \cite{Bechhoefer1989,Ignes-Mullol2000,Gomes2001,Oswald1991,Simon1990,Mesquita1996} have observed characteristics length scales on the order of tens and hundreds of microns.  To circumvent this, \added[NMA]{a destabilizing temperature gradient is imposed} with both temperature boundary conditions (see Fig. \ref{fig:compdomain}, boundary conditions 1 and 4) below the bulk transition temperature for the material parameters used.  This allows for the observation of morphological instabilities at wavelengths accessible in the domain sizes.}{Note that both temperature boundary conditions (see Fig. \ref{fig:compdomain}, boundary conditions 1 and 4) are below the bulk transition temperature for the material parameters used.  This allows for the observation of morphological instabilities at wavelengths accessible in the domain sizes that are within computational limitations.}

For symmetry boundary conditions of $\bm{Q}$, vector symmetry considerations result in the the following invariants \cite{Abukhdeir2008b}:
\begin{equation}\label{eqn:symmetry_bcs}
\frac{\partial Q_{xx}}{\partial x_i} = 0;\frac{\partial Q_{yy}}{\partial x_i} = 0;Q_{xy} = Q_{yx} =0;\frac{\partial T}{\partial x_i}= 0
\end{equation}
where $x_i$ \replaced[NMA]{is the coordinate associated with the basis vector normal to the symmetry axis}{depends on the orientation of the vector normal to the boundary}.

A reference system moving with velocity $v$, in the $x$ direction and equal to the negative of the velocity of the moving isotropic/nematic interface, was used. This maintains a static interface position, where time derivatives are replaced by material derivatives:

\begin{equation}
\frac{{\partial }}{{\partial t}} \rightarrow \frac{{\partial }}{{\partial t}}-v\frac{{\partial }}{{\partial y}}\
\end{equation}

\replaced[NMA]{Initial conditions resulting in an initially flat unperturbed homeotropic interface were achieved using the Heaviside step function}{Simulations with an initially flat unperturbed homeotropic interface were performed.  The Heaviside step function was used to generate the initial condition for $\bm{Q}$}:
\begin{equation}\label{init_unperturbed_Q}
\bm{Q}=\bm{Q}_0 He(y-y_0)
\end{equation}
where $\bm{Q}_0$ is the bulk Q-tensor value corresponding to the interface temperature, $He()$ is the Heaviside step function, and $y_0$ is the initial position of the interface.  \replaced[NMA]{Simulations were run to determine steady-state, or stationary, interface profiles.  Using these computed profiles a perturbation condition was applied using the following transformation}{These simulations were run until a stationary state was reached.  Then, simulations for perturbed interfaces were run, using the following initial conditions}:
\begin{equation}\label{initial_perturbed_Q}
\bm{Q}(x)=\bm{Q}_{stat}\left(y + a\cos \left(\frac{\pi}{\lambda_0} x \right)e^{-\delta y} \right) 
\end{equation}
where $\bm{Q}_{stat}(y)$ is the stationary \added[NMA]{(computed)} profile of $\bm{Q}$, $a$ is the initial amplitude of the sinusoidal perturbation, $\lambda_0$ the initial \added[NMA]{half-}wavelength, and the coefficient $\delta$ \replaced[NMA]{is the decay factor.  The characteristic decay length, $\delta^{-1}$, was chosen to equal $200nm$ so that the perturbation decays rapidly in the bulk relative to initial length scale of the nematic domain (on the order of microns).}{ in the exponential is chosen so that the perturbation vanishes rapidly in the bulk nematic phase.}

The initial condition Eq. (\ref{initial_perturbed_Q}) represents a flat stationary interface with a small sinusoidal perturbation.  A half wavelength was resolved in the simulation domain, where symmetry was used to capture a periodicity and the simulation domain size $l$ was varied, to represent different wavelengths.  Due to the limitations of the finite-element software used (Comsol Multiphysics), a static mesh was iteratively determined for each simulation where mesh density ranged from a maximum of approximately one second-order Lagrange element per $4nm^2$ to the minimum of 1 element per $0.5 \mu m^2$.  Convergence, mesh independence, and accuracy was implemented using standard numerical procedures.  Validation of the numerical results were established using homogeneous states.

\section{Results and Discussion} \label{sec:resdisc}

\subsection{Linear regime: shape instability} \label{subsec:linear_analysis}

A sharp interface model for the isotropic/nematic mesophase front growth is presented in the Appendix.  This derivation differs from standard sharp interface models \cite{Flemmings1974,Narayan2002} in that it:
\begin{itemize}
\item accounts for different thermal conductivities of both phases.
\item there is no assumption that the ordered phase is isothermal
\item uses an interfacial nematodynamic model \cite{Wincure2006} for the velocity of the interface as a function of temperature (and not the typically used Gibbs-Thompson relation \cite{Flemmings1974}).  
\end{itemize}
The final expression derived from the simple sharp-interface model (refer to the appendix) for the dispersion diagrams, in the low amplitude linear regime is:
\begin{eqnarray}\label{eqn:disp_rel}
0 &=& \left[ {\frac{d \left(f^n \beta^{-1}\right) }{dT}}  \right]^{ - 1} \sigma  - \frac{v} {\alpha }C_n  + \frac{\gamma }{2}{\left[ {\frac{d \left(f^n \beta^{-1}\right) }{dT}}  \right]^{ - 1} \kappa ^2 }  \nonumber\\ 
& +& \frac{v + s_i  }{ s_n + s_i }\left( \frac{v}{\alpha_n }C^n  - \frac{v}{\alpha_i }C^i \right)  \nonumber\\ 
&+& \frac{2}{s_n + s_i} \left( \sigma \frac{L}{\rho Cp} - \alpha _n \left( {\frac{v} {{\alpha _n }}} \right)^2 C^n  + \alpha _i \left( {\frac{v} {{\alpha _i }}} \right)^2 C^i \right) \nonumber\\
s_j& =& \sqrt {v^2  + 4\alpha _j \left( {\kappa ^2  + \sigma } \right)},j=i,n \nonumber\\
C^n& =&\frac{T_1 - T_I}{e^{\frac{v}{\alpha ^n}}-1} \nonumber\\
C^i&=&\frac{T_2 - T_I}{e^{-\frac{v}{\alpha ^i}}-1}
\end{eqnarray}
where $\kappa=2\pi /\lambda$ is the wave vector, $\lambda$ is the wavelength, $\sigma$ is the growth coefficient, $f^n$ is the bulk nematic free energy at the interface temperature, $\beta$ is a surface viscosity (defined in the appendix), $L$ is the heat of phase transition, $\alpha = k/\rho Cp$ is the thermal diffusivity,  $v$ is the interface velocity, $\gamma$ is the surface tension.  The superscripts \textit{n} and \textit{i} refer to the nematic and isotropic phases, respectively. 

\begin{figure}[htp]
\begin{center}
\includegraphics[width=3in]{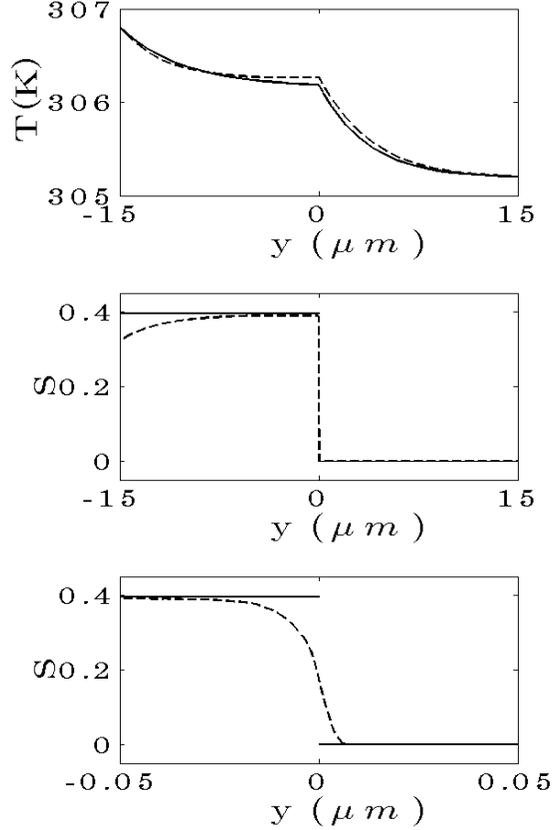} 
\end{center}
\caption{\added[NMA]{Sample order parameter and temperature profiles for initial flat interfaces from the sharp-interface model (solid lines) and stationary simulation results (dotted lines) a) (top) temperature versus y-coordinate over full domain b) (middle) uniaxial nematic scalar order parameter versus y-coordinate over full domain c) uniaxial nematic scalar order parameter versus y-coordinate magnified in the interface region.} \label{fig:modelcomp}}
\end{figure}

The sharp-interface model used to derive Eq. (\ref{eqn:disp_rel}) uses an important approximation: the value of the order parameter is assumed uniform in the nematic phase.  This approximation, although convenient for the derivation of an analytical solution, is not physically realistic.  The order parameter in the bulk nematic phase changes due to its dependence on temperature.  This non-uniformity in the order parameter, observed in the simulations, has an effect on the temperature profile which is not accounted for in the sharp interface model.  This effects results from the presence of temperature gradients which is accounted for in the energy balance (see term 3 of Eq. (\ref{eq:energy_balance})).\added[NMA]{  Sample order parameter and temperature profiles are shown in Figure \ref{fig:modelcomp} for both the sharp-interface model and the non-isothermal Landau-de Gennes model to elucidate this point}.
 
Simulations were performed using different initial wavelengths, under two different temperature gradients, where the initial perturbation amplitude used was much less than the wavelength.  The material properties and individual simulation parameters are given in Tables \ref{tab:parameters} and \ref{tab:sim_parameters}.  The boundary temperatures were selected so that the initially imposed interfacial temperature was equal for both gradients.  The perturbation amplitude, defined as the difference of the vertical positions of the interface at the symmetry axes (see Fig. \ref{fig:compdomain}) was obtained from the simulations as a function of time.  The position of the interface was \replaced[NMA]{determined from the contour at which the uniaxial nematic scalar order parameter equaled $S_i$}{defined as}:
\begin{equation}
S_i = \frac{S_b(T_i)}{2}
\end{equation}
where $S_b$ bulk value of the uniaxial nematic order parameter that corresponds to the initial interface temperature ($T_i$)\deleted[NMA]{and $S_i$ is the value of the uniaxial nematic order parameter that is assumed to define the interface}.  The lack of biaxiality at the interface on the symmetry axes was verified a posteriori.  Finally, the amplitude versus time was fitted with an exponential and the growth coefficient ($\sigma$), was obtained:
\begin{equation}
A(t)=A_0 exp \left(\sigma t\right)
\end{equation}

\begin{figure}[htp]
\begin{center}
\includegraphics[width=3in]{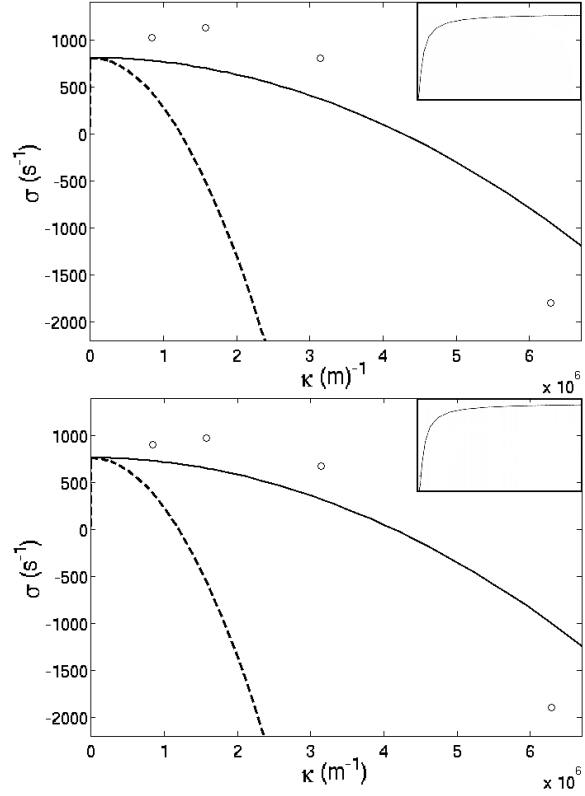} 
\end{center}
\caption{Dispersion diagram results directly from simulation (circles), using the analytical solution (dashed line, calculated using surface energy obtained from the definition in ref \cite{Wincure2006}, $4.5\times 10^{-4} \frac{J}{m^2}$), and using the analytical solution (full line, dispersion diagram calculated using surface energy = $3.0\times 10^{-5} \frac{J}{M^2}$ best fit from simulation).  \added[NMA]{Insets in the upper right corner of the plots show initial regime (low $\kappa$) are shown for reference}, a) (top) results corresponding to simulations $1a-1d$, b) (bottom) results corresponding to simulations $2a-2d$. \label{fig:dispdiag}}
\end{figure}

\begin{table}
\caption{Material properties for 5CB used in simulation \cite{Wincure2006,Pestov2003}. \label{tab:parameters}}
\begin{ruledtabular}
\begin{tabular}{|c|c|c|}
Parameter & Value & Units\\
\hline
$\mu$&$0.084$&$N \deleted[NMA]{\cdot} s/m^2$\\
$T_{NI}$&$307.2 $&$K$\\
$a_0$&$1.4\times 10^5 $&$J/m^3·K$\\
$b$&$1.8\times 10^7 $&$J/m^3$\\
$c$&$3.6\times 10^6 $&$J/m^3$\\
$l_1$&$3.0\times 10^{-12} $&$J/m$\\
$l_2$&$3.1\times 10^{-12} $&$J/m$\\
$l_3$&$0.0\times 10^{-12} $&$J/m$\\
$k_{\parallel}$&$0.2009 $&$W/m·K$\\
$k_{\perp}$&$0.1364 $&$W/m·K$\\
$C_p$&$1800 $&$J/kg·K$\\
$\rho$&$1000 $&$kg/m^3$
\end{tabular}
\end{ruledtabular}
\end{table}

\begin{table*}
\caption{Simulation parameters and growth coefficient ($\sigma$) results. \label{tab:sim_parameters}}
\begin{ruledtabular}
\begin{tabular}{|l|l|l|l|l|l|l|l|l|}
$\#$&$B (\mu m)$&$T_1 (K)$&$T_2 (K)$ & $\lambda_0 (\mu m)$ & $v (m/s)$ & $a (\mu m)$ & $T_i (K)$ & $\sigma (\times 10^{-11} s^{-1})$ \\
\hline
1a & 15 & 307.4 & 305.0 & 1 & 0.01955 & 0.06 & 306.27& -9.23\\
1b & 15 & 307.4 & 305.0 & 2 & 0.0195 & 0.06 & 306.27 & 4.134\\
1c & 15 & 307.4 & 305.0 & 4 & 0.0195 & 0.06 & 306.27 & 5.784\\
1d & 15 & 307.4 & 305.0 & 7.5 & 0.0196 & 0.09 & 306.27 & 5.282\\
2a & 15 & 306.8 & 305.4 & 1 & 0.0202 & 0.06 & 306.23 & -9.744\\
2b & 15 & 306.8 & 305.4 & 2 & 0.02035 & 0.06 & 306.23 & 3.467\\
2c & 15 & 306.8 & 305.4 & 4 & 0.02045 & 0.06 & 306.23 & 4.979\\
2d & 15 & 306.8 & 305.4 & 7.5 & 0.02005 & 0.09 & 306.23 & 4.615\\
\end{tabular}
\end{ruledtabular}
\end{table*}

The results are shown in Fig. \ref{fig:dispdiag}, where the growth coefficient obtained from simulations for different wavelengths in the small-amplitude regime are compared with those obtained from the sharp-interface model using two different values of the surface energy: a value from previous work \cite{Wincure2007} and the values estimated from both sets of simulations. The sharp interface mode predicts a sharp decrease in the growth coefficient in the vicinity of $\kappa=0$ which is shown in the insets of Fig. \ref{fig:dispdiag}.

\begin{figure*}[htp]
\begin{center}
\includegraphics[width=6in]{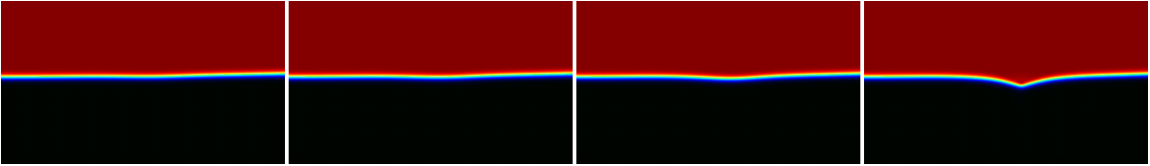} 
\end{center}
\caption{\added[NMA]{(Color online)} initially homeotropic transition from the linear regime i) $t=156.0 \mu s$ ii) $t=175.5 \mu s$ iii) $t=195.0 \mu s$ iv) $t=214.5 \mu s$; the surface corresponds to the scalar uniaxial nematic order parameter (red/black corresponds to isotropic/nematic), and the horizontal length scale of the figure is $1 \mu m$. \label{fig:homeo1}}.
\end{figure*}

As previously explained (see Sec. \ref{sec:intro}), a value of $\kappa$ exists for which the growth coefficient is maximum.  For large values of $\kappa$ (low wavelength), curvature effects are predominant and thus the growth coefficient is negative (the perturbation shrinks). At intermediate values, the growth coefficient is positive because of the \added[NMA]{increased destabilizing} effect of the \replaced[NMA]{temperature gradient at the interface}{temperature gradient}. As the value of $\kappa$ approaches zero, when the wavelength becomes much larger than the characteristic length of the temperature profile, the interface behaves as a flat interface and the problem becomes essentially one-dimensional. When the external length $B$ (see Fig. \ref{fig:compdomain}) is larger than the characteristic thermal length (as in the present case), the temperature at the interface in this one-dimensional problem depends only slightly on the position of the interface.  Thus, recalling Eq. \ref{eqn:velocity_interface}, the growth coefficient will be small.  As $B$ approaches $\infty$ (as usually assumed in theoretical studies), the temperature at the interface becomes independent of position and the growth coefficient is zero. 

The parameters involved in the calculation of the sharp-interface dispersion diagrams have been calculated from simulation results, as in ref \cite{Wincure2007} (see Fig. \ref{fig:dispdiag}).  The calculated dispersion diagram with these parameters reproduces well the maximum value of the growth coefficient, as compared with simulations, but the curve resulting from the surface tension definition from ref \cite{Wincure2007} is more narrow than that determined from direct numerical simulation.

\subsection{Non-linear regime: structural dynamics} \label{subsec:nonlinear_analysis}

The shape instability due to the presence of temperature gradients was not the only interfacial instability that was observed.  A process of defect formation, shedding, and the growth of a disoriented planar domain was also observed following the initial linear regime.  It is important to note that results presented in the previous section were in the transient linear instability regime.  The nonlinear regime instabilities presented in this section were observed to follow the initially linear instabilities

\begin{figure*}[htp]
\begin{center}
\includegraphics[width=6in]{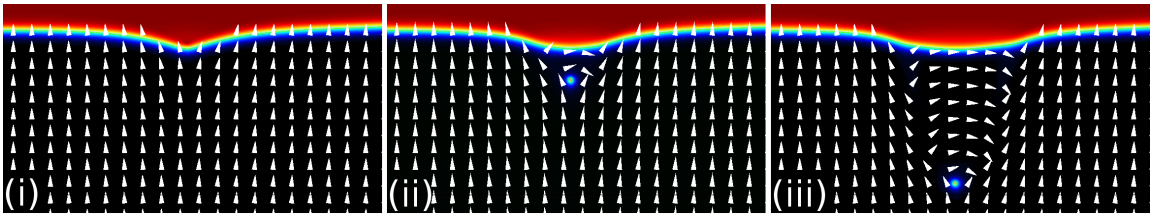} 
\end{center}
\caption{\added[NMA]{(Color online)} initially homeotropic defect formation i) $t=214.5 \mu s$ ii) $t=219.4 \mu s$ iii) $t=229.1 \mu s$; the surface corresponds to the scalar uniaxial nematic order parameter (red/black corresponds to isotropic/nematic), arrows correspond to the uniaxial nematic director (should be considered headless), and horizontal length scale is $0.65 \mu m$\label{fig:homeo2}}
\end{figure*}

\begin{figure*}[htp]
\begin{center}
\includegraphics[height=1.65in]{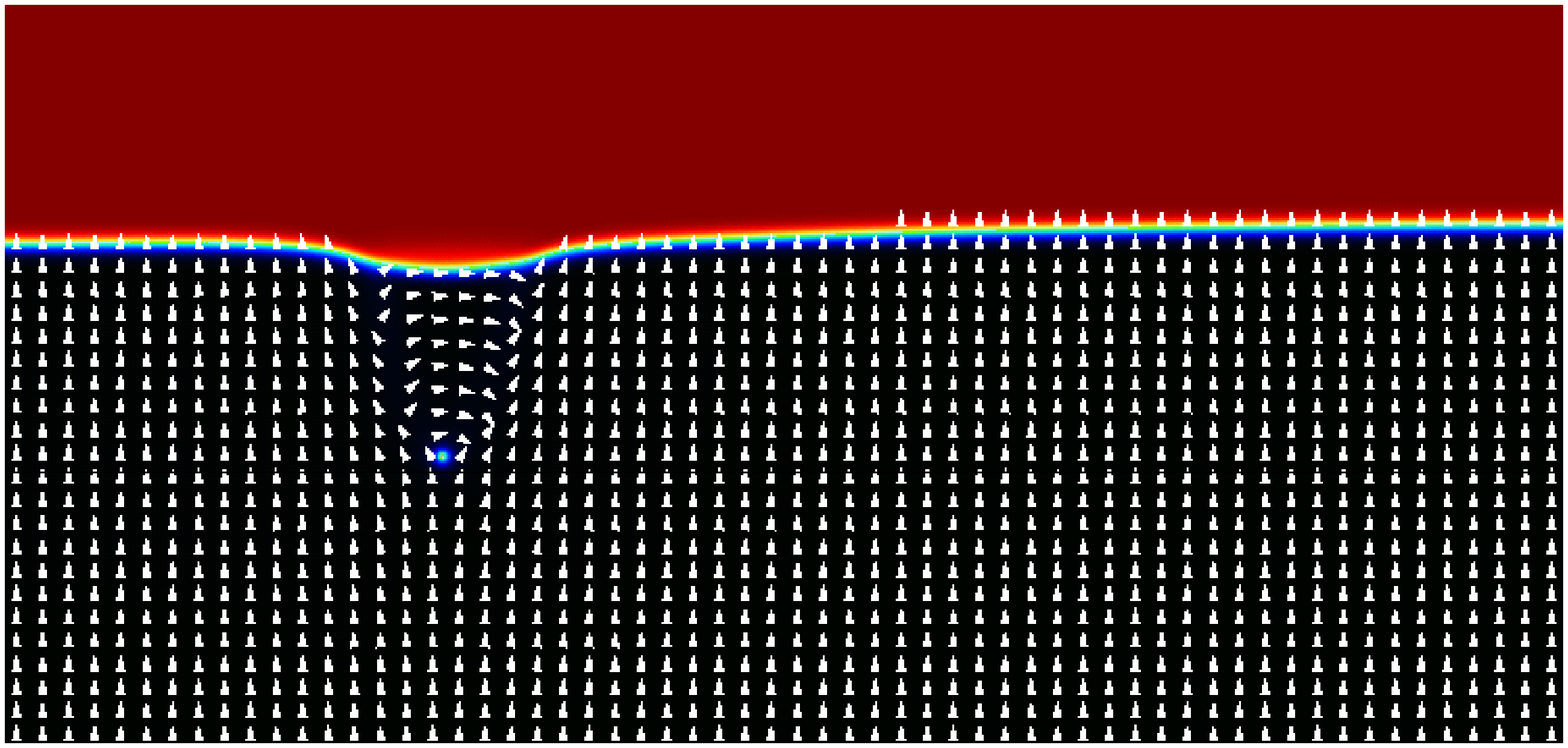} 
\includegraphics[height=1.65in]{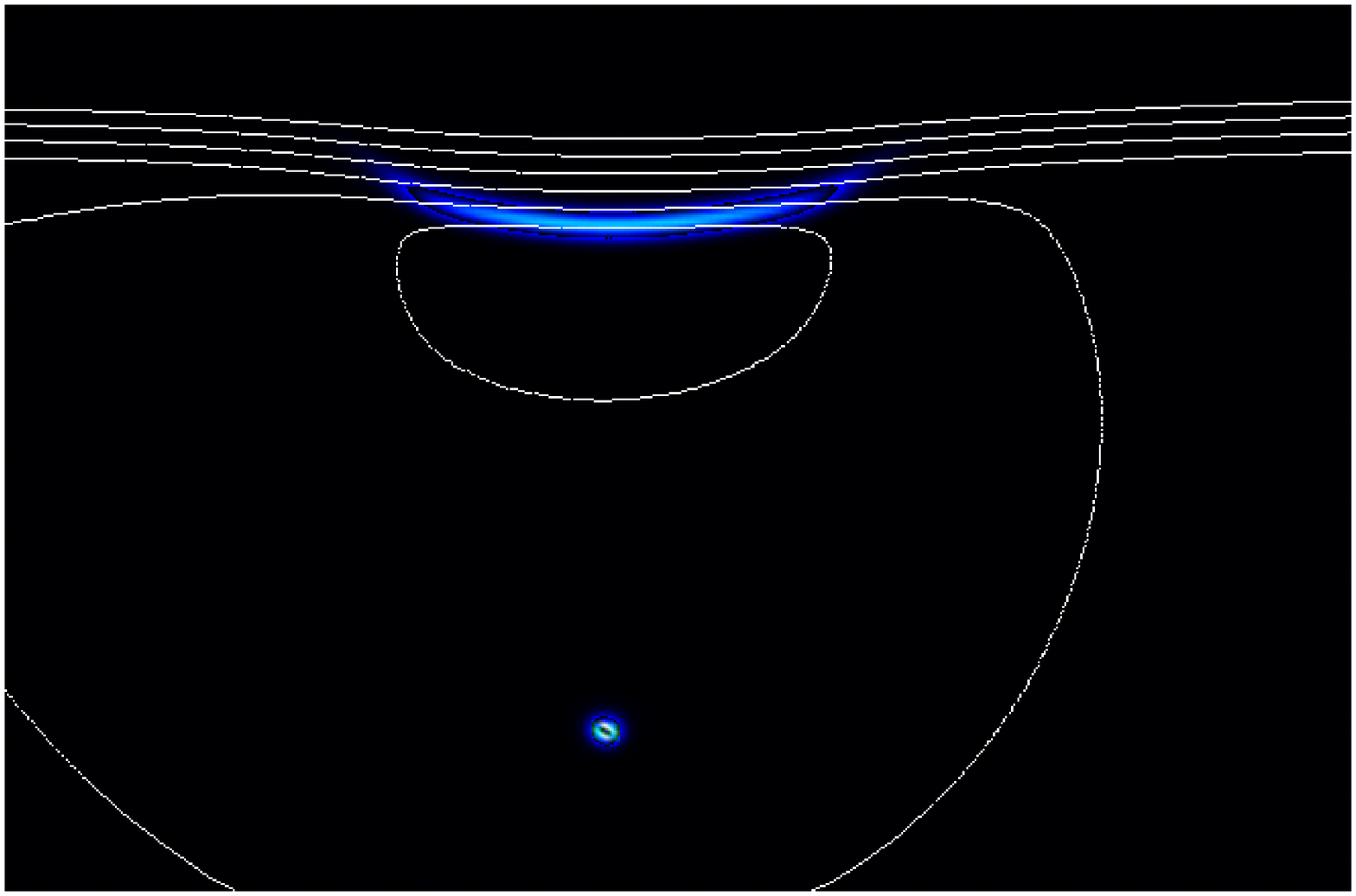} 
\end{center}
\caption{\added[NMA]{(Color online)} $t = 229.1 \mu s$ (corresponds to Fig. \ref{fig:homeo1}(iii) a) (left) view of full computational domain where the surface corresponds to the scalar uniaxial nematic order parameter (red/black corresponds to isotropic/nematic), arrows correspond to the uniaxial nematic director (should be considered headless), and the horizontal length scale is $2 \mu m$ b) (right) magnified view of the planar interface with the surface corresponding to $\beta^2$ (see Eq. (\ref{eqn:biax})) and the contours corresponding to temperature (the minimum/maximum/increment is $306.285K/306.295K/0.002K$) the horizontal length scale is $0.6 \mu m$. \label{fig:homeo3}}
\end{figure*}

These nonlinear regime instabilities are shown in Figs. \ref{fig:homeo1}-\ref{fig:homeo3} which correspond to simulation $2c$ (see Table \ref{tab:sim_parameters}).  Referring to Fig. \ref{fig:homeo1}, a cusp is formed at the interface as the instability transitions from the linear regime, discussed in the Sec. \ref{subsec:linear_analysis}, to the nonlinear regime.  As the cusp sharpens, it forms a $+\frac{1}{2}$ disclination which is then shed into the bulk, similar to the defect shedding mechanism found by Wincure and Rey in growing initially homogeneous nematic spherulites under isothermal conditions \cite{Wincure2007a}.  As the defect sheds, a planar anchored interfacial regime nucleates and grows with the moving front.  Fig. \ref{fig:homeo3}a shows the full view of the computational domain with the director profile.  Fig. \ref{fig:homeo3}b shows the temperature profile and biaxial nature of the $+\frac{1}{2}$ disclination and planar isotropic/nematic interface, computed as follows \cite{Kaiser1992,Kralj1999}:
\begin{equation}\label{eqn:biax}
\beta^2 = 1-6\frac{\left[\left(\bm{Q}\cdot\bm{Q}\right) : \bm{Q}\right]^2}{\left(\bm{Q} : \bm{Q}\right)^3}
\end{equation}
ranging from ranging from fully uniaxial $\beta^2 = 0$ to fully biaxial $\beta^2=1$. 

\begin{figure}[htp]
\begin{center}
\includegraphics[width=3in]{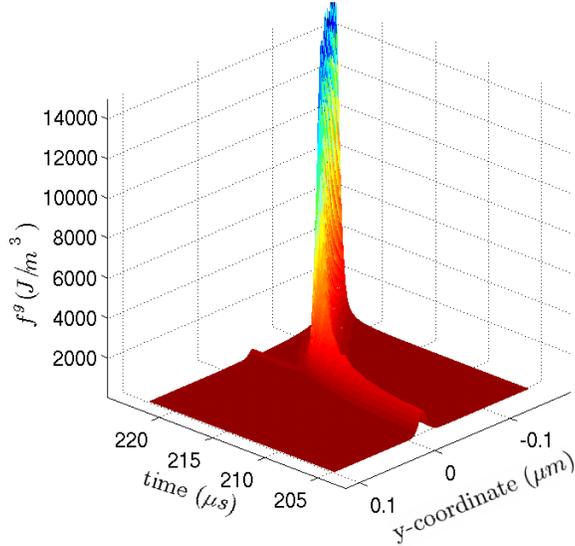} 
\end{center}
\caption{\added[NMA]{(Color online)} Surface plot of the gradient free energy density (refer to Eq. (\ref{eq:free_energy_heterogeneous})) as a function of time and y-coordinate ($\mu m$); features of the energy surface include the initial homeotropically oriented interface, defect shedding (the high energy peak), and the relatively lower energy planar oriented interface. \label{fig:energy_plots}}
\end{figure}

\added[NMA]{Biaxiality has a key role in this process, where the transition to an inherently biaxial planar interface from a uniaxial homeotropic interface is achieved through the shedding of a biaxial disclination defect.  Exhaustive past work has addressed these interfacial texturing processes for isothermal isotropic/nematic interfaces, see refs. \cite{Wincure2006,Wincure2007} for a full treatment.}  This defect formation, shedding, and planar interfacial growth phenomena are driven by planar anchoring having a lower energy than that of a homeotropic anchoring.  This is shown in Fig. \ref{fig:energy_plots}, which shows the gradient free energy density $f_g$ (terms 4-5 Eq. (\ref{eq:free_energy_heterogeneous})):
\begin{equation}
f_g = \frac{1}{2} l_1 (\bm{\nabla} \bm{Q} \vdots \bm{\nabla} \bm{Q} ) + \frac{1}{2} l_2 \left( \nabla \cdot \bm{Q} \right) \cdot \left( \nabla \cdot \bm{Q} \right)
\end{equation}
across the interface at the horizontal position at which the defect forms, as a function of the position (across the interface) and time.  Before the defect formation, the interfacial anchoring is homogeneous in the high-energy homeotropic state.  As the disclination forms, the interfacial anchoring becomes heterogeneous where an energetically favorable planar regime forms trailing the shed defect.  In the vicinity of the disclination defect the free energy is greatly increased; it is important to note that the total free energy of the system is minimized in that the energetically favorable interfacial region of planar anchoring grows, decreasing the total free energy.

In order to exhaustively confirm that this instability results from the difference in anchoring energies, a simulation was performed with $l_2=0$.  This is based upon the notion that for an interface, neglecting curvature and biaxiality, the surface energy is predicted to be \cite{Yokoyama1997}:
\begin{equation} \label{eqn:int_tens}
\gamma=\frac{b^3\sqrt{3l_1+l_2/2+3l_2(\bm{n\cdot k})^2/2}}{486c^{5/3}}
\end{equation}
where $\bm{k}$ is the unit vector normal to the interface.  Equation (\ref{eqn:int_tens}) shows that the sign of $l_2$ determines which orientation has the lower surface energy; if it is positive it will planar ($\bm{n}\perp\bm{k}$), and if it is negative, homeotropic ($\bm{n}=\bm{k}$).  When $l_2=0$, neither of the interfacial anchorings are preferred. Thus if the surface free energy is the driving force for defect formation; no defect would be expected to form when $l_2=0$ which was confirmed via this simulation where, for long times, no defect formation  was observed.

\section{Conclusions} \label{sec:conc}

In this work, two-dimensional multi-scale simulation of the isotropic/nematic mesophase was used to study both linear and nonlinear morphological instabilities under directional growth.  The major contributions of this work are:
\begin{itemize}
\item an extended sharp-interface model was derived (refer to Sec. \ref{subsec:linear_analysis} and the appendix) and applied to linear morphological instabilities of the isotropic/nematic mesophase transition.
\item results from comparing the sharp-interface model and the two-dimensional simulation of the tensorial Landau-de Gennes model (refer to Sec. \ref{subsec:ord_model}) identify the limitations of sharp-interface models to predict the maximum value of the growth coefficient (see Fig. \ref{fig:dispdiag}).  It is shown that this type of model fails to adequately capture the time-dependent morphological instability growth compared to full two-dimensional simulation of a non-isothermal Landau-de Gennes model \cite{Abukhdeir2008b}.
\item nonlinear instabilities were studied for an initially homeotropic isotropic/nematic interface, where the phenomenological model predicts the planar isotropic/nematic interface being more stable.  Disclination shedding at the interface, similar to experimentally \cite{Bechhoefer1996} and theoretically observed phenomena \cite{Wincure2007a}, and the formation/growth of bulk texturing were found to result (see Figs. \ref{fig:homeo1}-\ref{fig:energy_plots}).
\end{itemize}
While these results confirm the validity of sharp-interface models for the well-studied problem of linear morphological instabilities, a substantial conclusion can be drawn from simulation results in the nonlinear regime.  That conclusion being that, using the full tensorial Landau-de Gennes model for the isotropic/nematic phase transition, extended to the non-isothermal case \cite{Abukhdeir2008b}, nonlinear morphological instabilities can be accessed that are both inaccessible using more simple theoretical approaches and difficult to characterize experimentally.

Furthermore, three-dimensional simulation using the tensorial Landau-de Gennes model \cite{Abukhdeir2008b}, based upon the current results, will enable the study of geometries that are more directly representative of experimentally conditions.  For example, some of the physics that can be accessed include surface anchoring, meniscus formation \cite{Oswald1991}, three-dimensional textures/defects et al.

\begin{acknowledgments}
 This research was supported in part by the National Research Council of Argentina (ERS) and the Natural Science and Engineering Research Council of Canada (NMA and ADR).
\end{acknowledgments}

\appendix
\section{Sharp interface model for linear stability of a planar interface growing in an external temperature gradient} \label{sec:app1}

The objective of this appendix is to derive an analytical expression for the dispersion diagram, considering a sharp-interface model.  The energy balance for the bulk nematic and isotropic phases, in a frame moving with velocity $v$ are:

\begin{equation} \label{eqn:conduction_nematic}
\frac{{\partial T^n }}{{\partial t}} - v\frac{{\partial T^n }} {{\partial y}} = \alpha ^n\left( {\frac{{\partial ^2 T^n }} {{\partial y^2 }} + \frac{{\partial ^2 T^n }} {{\partial x^2 }}} \right)
\end{equation}

\begin{equation} \label{eqn:conduction_isotropic}
\frac{{\partial T^i}}{{\partial t}} - v\frac{{\partial T^i}}{{\partial y}} = \alpha ^i\left( {\frac{{\partial ^2 T^i}}{{\partial x^2}} + \frac{{\partial ^2 T^i}}{{\partial y^2}}} \right)
\end{equation}
where $\alpha$ is the thermal diffusivity. The superscripts $i$ and $n$ refer to the isotropic and nematic phase, respectively. We are assuming that the order parameter is uniform in the whole nematic phase.

The boundary conditions are:
\begin{equation} \label{eqn:borderdirichlet}
\left.{T}\right|_{y=-B}=T_1 \qquad \qquad \left.{T}\right|_{y=B}=T_2
\end{equation}
and, additionally, at the interface $y=h(x)$ the temperature must be continuous and the heat released by the phase transition must equal the net heat flux.

For a steady-state flat interface (base solution), we take the velocity of the moving reference system $v$ as the velocity of the interface, and place the interface at the origin, so $h(x)=0$. The temperature is only a function of $y$, and the solution to Eqs. (\ref{eqn:conduction_nematic}) and (\ref{eqn:conduction_isotropic}) is:
        
\begin{eqnarray} \label{eqn:base_solution}
T^n=C^n \left\{\exp\left(-\frac{v}{\alpha _n}y\right)-1\right\}+T_I \nonumber\\
T^i=C^i \left\{\exp\left(-\frac{v}{\alpha _i}y\right)-1\right\}+T_I
\end{eqnarray}
where $C^n$ and $C^i$ are integration constants, and $T_I$ is the temperature at the interface\replaced[NMA]{.  The boundary conditions Eq. \ref{eqn:borderdirichlet} and previously mentioned interface conditions are then used to find the particular solution from the general solution Eq. \ref{eqn:base_solution}}{, which is obtained from the boundary conditions}:
\begin{eqnarray} \label{constants} 
T_I &=& \{T_1 \left(f(-v)-1 \right) - T_2 \left(f(v)- 1 \right) \nonumber\\
&&+ \frac{L}{\rho Cp}\left( f(v)  - 1 \right)\left( f(-v)- 1 \right)\}/\{f(-v)  - f(v)\} \nonumber \\
&&f_i(v)= \exp\left(\frac{v} {\alpha _i} B\right) ; f_n(v)= \exp\left(\frac{v} {\alpha _n} B\right) \nonumber\\
&&C^n =\frac{T_1 - T_I}{\exp \left(\frac{v}{\alpha ^n}\right)-1}  ;C^i=\frac{T_2 - T_I}{\exp \left(-\frac{v}{\alpha ^i}\right)-1}
\end{eqnarray}
Now a perturbation to the base solution is considered:
\begin{eqnarray} \label{eqn:perturbed_T}
T^i=T^i_b + T^{*i} \nonumber\\
T^n=T^n_b + T^{*n}
\end{eqnarray}
where the $T^{*i}$ and $T^{*n}$ are the perturbations, and the subscript $b$ are the base solutions.  Replacing Eq. (\ref{eqn:perturbed_T}) in Eqs. (\ref{eqn:conduction_nematic}) and (\ref{eqn:conduction_isotropic}):
\begin{eqnarray} \label{eqn:conduction_perturbation}
\frac{{\partial T^{*n} }}{{\partial t}} - v\frac{{\partial T^{*n} }} {{\partial y}} &=& \alpha ^n\left( {\frac{{\partial ^2 T^{*n} }} {{\partial y^2 }} + \frac{{\partial ^2 T^{*n} }} {{\partial x^2 }}} \right) \\
\frac{{\partial T^{*i} }}{{\partial t}} - v\frac{{\partial T^{*i} }} {{\partial y}} &=& \alpha ^i\left( {\frac{{\partial ^2 T^{*i} }} {{\partial y^2 }} + \frac{{\partial ^2 T^{*i} }}
{{\partial x^2 }}} \right)
\end{eqnarray}
As long the perturbations are small and do not reach the borders of the simulation domain, boundary conditions at infinity can be used:\begin{equation} \label{eqn:border_perturbation}
\left. {T^{*n}} \right|_{y=-\infty} =0 \qquad \qquad   \left. {T^{*i}} \right|_{y=\infty} =0
\end{equation}
Writing the boundary conditions at the interface, considering that now $h = h(x)$ and linearizing, the following boundary conditions are obtained, where the temperatures and their derivatives are evaluated at $y=0$:
\begin{eqnarray}\label{eqn:interface_conditions_perturbation} 
\frac{\partial T^n_b}{\partial y}h + T^{*n} &=& \frac{\partial T^i_b}{\partial y}h + T^{*i} \\ 
k_n \! \left( \frac{\partial T^{*n}}{\partial y} + \frac{\partial^2 T^{*n}}{\partial^2 y} \right) &=&k_i  \left( \frac{\partial T^{*i}}{\partial y} + \frac{\partial^2 T^{*i}}{\partial^2 y} \right)\nonumber\\
&& + \frac{dh}{dt}L
\end{eqnarray}
The velocity of the nematic-isotropic interface, can be calculated from a nematodynamic interfacial model \cite{Wincure2006}:
\begin{eqnarray}\label{eqn:interface_velocity}
\beta w = \left( \bm{L} + \nabla_s  \cdot \bm{T}_s  \right) \cdot \bm{k} + \mu \bm{Q}^s :\frac{d \bm{Q}^s }{dt}\\
\beta=\int_{\Delta_N}^{\Delta_I} \frac{\partial \bm{Q}}{\partial \lambda}:\frac{\partial \bm{Q}}{\partial \lambda}h d\lambda
\end{eqnarray}
where $\beta$ is the interfacial viscosity, $\bm{L}$ is the temperature dependent net stress load (which reduces to the free energy difference between the nematic and the isotropic phase), $\nabla_s \cdot \bm{T}_s $ is the capillary force (neglecting anisotropy in surface tension, it is the product of the surface tension and the bidimensional curvature of the interface), and the last term is the change in the value of the order parameter at the interface (assumed to be 0).  If the perturbation is small, Eq. \ref{eqn:interface_velocity} reduces to:
\begin{equation}\label{eqn:interface_velocity_simplified}
\beta w = -f^n + \frac{\gamma}{2} \frac{d^2h}{dx^2}
\end{equation}
where $f^n$ is the bulk nematic free energy, and $\gamma$ is the surface tension.  The surface tension, neglecting curvature, is found to be:
\begin{equation}
\gamma=\frac{1}{2}\bm{I_S}:\int_{\Delta_N}^{\Delta_I}\left( {f^n\bm{I_S} -\frac{\partial f^n}{\partial \nabla \bm{Q}}:(\nabla \bm{Q})^T} \right) d\lambda
\end{equation}
where $\bm{I_S}$ is the surface $2\times2$ identity matrix.
After linearization, considering that for small perturbations the normal to the interface is approximately the $y$ direction, the change in the velocity with respect to the base solution (the velocity of the interface in the moving frame) is: 
\begin{equation}\label{eqn:dhdt}
\frac{dh}{dt} =  - \frac{d\left( f^n \beta^{-1} \right)}{dT} \left[ \left. \frac{\partial T^n_b }{\partial y} \right|_{y = 0} h + \left.T^{*n}\right|_{y = 0}  \right] + \frac{\gamma}{2\beta} \frac{d^2 h} {dx^2}
 \end{equation}
where the derivative with respect to temperature is a total derivative and thus the equilibrium order parameter as a function of the temperature must be used in the free energy.
 
The solution, for a sinusoidal perturbation at the interface, is: 
\begin{eqnarray}\label{eqn:perturbation_solution}
T^{*i} &=& T^i_0 \exp(a_i y) \exp(\sigma t)\sin(\kappa x) \nonumber\\
T^{*n} &=& T^i_0 \exp(a_n y) \exp(\sigma t)\sin(\kappa x) \\
h &=& h_0 e^{\sigma t}\sin(\kappa x) \nonumber
\end{eqnarray}
replacing in Eq. (\ref{eqn:conduction_perturbation}) results in:
\begin{eqnarray}\label{eqn:a_sigma_kappa}
\sigma - v a_i = \alpha_i\left( {a_i^2-\kappa^2} \right) \nonumber\\
\sigma - v a_i = \alpha_n\left( {a_n^2-\kappa^2} \right)
\end{eqnarray}
 
The system formed by Eqs. (\ref{eqn:interface_conditions_perturbation}, \ref{eqn:dhdt}, \ref{eqn:perturbation_solution}, \ref{eqn:a_sigma_kappa}) is a homogeneous system.  In order to have a non trivial solution, the following dispersion relation must be satisfied: 
\begin{eqnarray}\label{eqn:dispersion_relation}
0&=& \left[ {\frac{d \left(f^n \beta^{-1}\right) }{dT}}  \right]^{ - 1} \sigma  - \frac{v} {\alpha }C_n  + \frac{\gamma }{2 \beta}{\left[ {\frac{d \left(f^n \beta^{-1}\right) }{dT}}  \right]^{ - 1} \kappa ^2 }  \nonumber\\ 
  &+& \left( \frac{v}{\alpha_n }C^n  - \frac{v}{\alpha_i }C^i \right) \left( v + s_i \right) \left( s_n + s_i \right)^{-1} \\ 
&+& 2 \left( \sigma \frac{L}{\rho Cp} - \alpha _n \left( {\frac{v} {{\alpha _n }}} \right)^2 C^n  + \alpha _i \left( {\frac{v} {{\alpha _i }}} \right)^2 C^i \right) \left( s_n + s_i \right)^{-1}\nonumber
\end{eqnarray}
where:
\begin{equation}
s_j = \sqrt {v^2  + 4\alpha _j \left( {\kappa ^2  + \sigma } \right)},j=i,n
\end{equation}

\listofchanges

\bibliographystyle{apsrev}
\bibliography{/home/nasser/nfs/references/references}

\begin{thebibliography}{45}
\expandafter\ifx\csname natexlab\endcsname\relax\def\natexlab#1{#1}\fi
\expandafter\ifx\csname bibnamefont\endcsname\relax
  \def\bibnamefont#1{#1}\fi
\expandafter\ifx\csname bibfnamefont\endcsname\relax
  \def\bibfnamefont#1{#1}\fi
\expandafter\ifx\csname citenamefont\endcsname\relax
  \def\citenamefont#1{#1}\fi
\expandafter\ifx\csname url\endcsname\relax
  \def\url#1{\texttt{#1}}\fi
\expandafter\ifx\csname urlprefix\endcsname\relax\def\urlprefix{URL }\fi
\providecommand{\bibinfo}[2]{#2}
\providecommand{\eprint}[2][]{\url{#2}}

\bibitem[{\citenamefont{Pelc\'e et~al.}(2004)\citenamefont{Pelc\'e, Brujic, and
  Costier}}]{Pelce2004}
\bibinfo{author}{\bibfnamefont{P.}~\bibnamefont{Pelc\'e}},
  \bibinfo{author}{\bibfnamefont{J.}~\bibnamefont{Brujic}}, \bibnamefont{and}
  \bibinfo{author}{\bibfnamefont{L.}~\bibnamefont{Costier}},
  \emph{\bibinfo{title}{New Visions on Form and Growth: Fingered Growth,
  Dendrites, and Flames}} (\bibinfo{publisher}{Oxford University Press},
  \bibinfo{year}{2004}).

\bibitem[{\citenamefont{Mullins and Sekerka}(1963)}]{Mullins1963}
\bibinfo{author}{\bibfnamefont{W.~W.} \bibnamefont{Mullins}} \bibnamefont{and}
  \bibinfo{author}{\bibfnamefont{R.~F.} \bibnamefont{Sekerka}},
  \bibinfo{journal}{Journal of Applied Physics} \textbf{\bibinfo{volume}{34}},
  \bibinfo{pages}{323} (\bibinfo{year}{1963}),
  \urlprefix\url{http://link.aip.org/link/?JAP/34/323/1}.

\bibitem[{\citenamefont{Mullins and Sekerka}(1964)}]{Mullins1964}
\bibinfo{author}{\bibfnamefont{W.~W.} \bibnamefont{Mullins}} \bibnamefont{and}
  \bibinfo{author}{\bibfnamefont{R.~F.} \bibnamefont{Sekerka}},
  \bibinfo{journal}{Journal of Applied Physics} \textbf{\bibinfo{volume}{35}},
  \bibinfo{pages}{444} (\bibinfo{year}{1964}),
  \urlprefix\url{http://link.aip.org/link/?JAP/35/444/1}.

\bibitem[{\citenamefont{Coriell et~al.}(1985)\citenamefont{Coriell, McFadden,
  and Sekerka}}]{Coriell1985}
\bibinfo{author}{\bibfnamefont{S.}~\bibnamefont{Coriell}},
  \bibinfo{author}{\bibfnamefont{G.}~\bibnamefont{McFadden}}, \bibnamefont{and}
  \bibinfo{author}{\bibfnamefont{R.}~\bibnamefont{Sekerka}},
  \bibinfo{journal}{Annual Reviews in Materials Science}
  \textbf{\bibinfo{volume}{15}}, \bibinfo{pages}{119} (\bibinfo{year}{1985}).

\bibitem[{\citenamefont{Wheeler et~al.}(1993)\citenamefont{Wheeler, Murray, and
  Schaefer}}]{Wheeler1993}
\bibinfo{author}{\bibfnamefont{A.}~\bibnamefont{Wheeler}},
  \bibinfo{author}{\bibfnamefont{B.}~\bibnamefont{Murray}}, \bibnamefont{and}
  \bibinfo{author}{\bibfnamefont{R.}~\bibnamefont{Schaefer}},
  \bibinfo{journal}{Physica D} \textbf{\bibinfo{volume}{66}},
  \bibinfo{pages}{243} (\bibinfo{year}{1993}).

\bibitem[{\citenamefont{Braun et~al.}(1994)\citenamefont{Braun, McFadden, and
  Coriell}}]{Braun1994}
\bibinfo{author}{\bibfnamefont{R.~J.} \bibnamefont{Braun}},
  \bibinfo{author}{\bibfnamefont{G.~B.} \bibnamefont{McFadden}},
  \bibnamefont{and} \bibinfo{author}{\bibfnamefont{S.~R.}
  \bibnamefont{Coriell}}, \bibinfo{journal}{Phys. Rev. E}
  \textbf{\bibinfo{volume}{49}}, \bibinfo{pages}{4336} (\bibinfo{year}{1994}).

\bibitem[{\citenamefont{Wheeler et~al.}(1995)\citenamefont{Wheeler, Ahmad,
  Boettinger, Braun, McFadden, and Murray}}]{Wheeler1995}
\bibinfo{author}{\bibfnamefont{A.}~\bibnamefont{Wheeler}},
  \bibinfo{author}{\bibfnamefont{N.}~\bibnamefont{Ahmad}},
  \bibinfo{author}{\bibfnamefont{W.}~\bibnamefont{Boettinger}},
  \bibinfo{author}{\bibfnamefont{R.}~\bibnamefont{Braun}},
  \bibinfo{author}{\bibfnamefont{G.}~\bibnamefont{McFadden}}, \bibnamefont{and}
  \bibinfo{author}{\bibfnamefont{B.}~\bibnamefont{Murray}},
  \bibinfo{journal}{Advances in Space Research} \textbf{\bibinfo{volume}{16}},
  \bibinfo{pages}{163} (\bibinfo{year}{1995}).

\bibitem[{\citenamefont{Sekerka}(2004)}]{Sekerka2004}
\bibinfo{author}{\bibfnamefont{R.~F.} \bibnamefont{Sekerka}},
  \bibinfo{journal}{Journal of Crystal Growth} \textbf{\bibinfo{volume}{264}},
  \bibinfo{pages}{530} (\bibinfo{year}{2004}),
  \urlprefix\url{http://www.sciencedirect.com/science/article/B6TJ6-4BKN2XH-3/%
2/b3cb98419339f0e2cc7874b1787a245a}.

\bibitem[{\citenamefont{Sekerka}(2005)}]{Sekerka2005}
\bibinfo{author}{\bibfnamefont{R.~F.} \bibnamefont{Sekerka}},
  \bibinfo{journal}{Crystal Research and Technology}
  \textbf{\bibinfo{volume}{40}}, \bibinfo{pages}{291} (\bibinfo{year}{2005}),
  \urlprefix\url{http://dx.doi.org/10.1002/crat.200410342}.

\bibitem[{\citenamefont{Xu et~al.}(2005)\citenamefont{Xu, Keawwattana, and
  Kyu}}]{Xu2005a}
\bibinfo{author}{\bibfnamefont{H.}~\bibnamefont{Xu}},
  \bibinfo{author}{\bibfnamefont{W.}~\bibnamefont{Keawwattana}},
  \bibnamefont{and} \bibinfo{author}{\bibfnamefont{T.}~\bibnamefont{Kyu}},
  \bibinfo{journal}{The Journal of Chemical Physics}
  \textbf{\bibinfo{volume}{123}}, \bibinfo{pages}{124908}
  (\bibinfo{year}{2005}).

\bibitem[{\citenamefont{Bechhoefer et~al.}(1989)\citenamefont{Bechhoefer,
  Simon, Libchaber, and Oswald}}]{Bechhoefer1989}
\bibinfo{author}{\bibfnamefont{J.}~\bibnamefont{Bechhoefer}},
  \bibinfo{author}{\bibfnamefont{A.~J.} \bibnamefont{Simon}},
  \bibinfo{author}{\bibfnamefont{A.}~\bibnamefont{Libchaber}},
  \bibnamefont{and} \bibinfo{author}{\bibfnamefont{P.}~\bibnamefont{Oswald}},
  \bibinfo{journal}{Phys. Rev. A} \textbf{\bibinfo{volume}{40}},
  \bibinfo{pages}{2042} (\bibinfo{year}{1989}).

\bibitem[{\citenamefont{Bechhoefer and Langer}(1995)}]{Bechhoefer1995}
\bibinfo{author}{\bibfnamefont{J.}~\bibnamefont{Bechhoefer}} \bibnamefont{and}
  \bibinfo{author}{\bibfnamefont{S.~A.} \bibnamefont{Langer}},
  \bibinfo{journal}{Phys. Rev. E} \textbf{\bibinfo{volume}{51}},
  \bibinfo{pages}{2356} (\bibinfo{year}{1995}).

\bibitem[{\citenamefont{Ign\'es-Mullol and Oswald}(2000)}]{Ignes-Mullol2000}
\bibinfo{author}{\bibfnamefont{J.}~\bibnamefont{Ign\'es-Mullol}}
  \bibnamefont{and} \bibinfo{author}{\bibfnamefont{P.}~\bibnamefont{Oswald}},
  \bibinfo{journal}{Phys. Rev. E} \textbf{\bibinfo{volume}{61}},
  \bibinfo{pages}{3969} (\bibinfo{year}{2000}).

\bibitem[{\citenamefont{Gomes et~al.}(2001)\citenamefont{Gomes, Falc\~ao, and
  Mesquita}}]{Gomes2001}
\bibinfo{author}{\bibfnamefont{O.~A.} \bibnamefont{Gomes}},
  \bibinfo{author}{\bibfnamefont{R.~C.} \bibnamefont{Falc\~ao}},
  \bibnamefont{and} \bibinfo{author}{\bibfnamefont{O.~N.}
  \bibnamefont{Mesquita}}, \bibinfo{journal}{Phys. Rev. Lett.}
  \textbf{\bibinfo{volume}{86}}, \bibinfo{pages}{2577} (\bibinfo{year}{2001}).

\bibitem[{\citenamefont{Wincure and Rey}(2006)}]{Wincure2006}
\bibinfo{author}{\bibfnamefont{B.}~\bibnamefont{Wincure}} \bibnamefont{and}
  \bibinfo{author}{\bibfnamefont{A.}~\bibnamefont{Rey}}, \bibinfo{journal}{The
  Journal of Chemical Physics} \textbf{\bibinfo{volume}{124}},
  \bibinfo{eid}{244902} (pages~\bibinfo{numpages}{13}) (\bibinfo{year}{2006}).

\bibitem[{\citenamefont{Wincure and Rey}(2007{\natexlab{a}})}]{Wincure2007a}
\bibinfo{author}{\bibfnamefont{B.}~\bibnamefont{Wincure}} \bibnamefont{and}
  \bibinfo{author}{\bibfnamefont{A.}~\bibnamefont{Rey}}, \bibinfo{journal}{Nano
  Letters} \textbf{\bibinfo{volume}{7}}, \bibinfo{pages}{1474}
  (\bibinfo{year}{2007}{\natexlab{a}}), ISSN \bibinfo{issn}{1530-6984}.

\bibitem[{\citenamefont{Oswald}(1991)}]{Oswald1991}
\bibinfo{author}{\bibfnamefont{P.}~\bibnamefont{Oswald}},
  \bibinfo{journal}{Journal de Physique II} \textbf{\bibinfo{volume}{1000}},
  \bibinfo{pages}{5} (\bibinfo{year}{1991}).

\bibitem[{\citenamefont{Simon and Libchaber}(1990)}]{Simon1990}
\bibinfo{author}{\bibfnamefont{A.~J.} \bibnamefont{Simon}} \bibnamefont{and}
  \bibinfo{author}{\bibfnamefont{A.}~\bibnamefont{Libchaber}},
  \bibinfo{journal}{Phys. Rev. A} \textbf{\bibinfo{volume}{41}},
  \bibinfo{pages}{7090} (\bibinfo{year}{1990}).

\bibitem[{\citenamefont{Mesquita et~al.}(1996)\citenamefont{Mesquita,
  Figueiredo, and Vidal}}]{Mesquita1996}
\bibinfo{author}{\bibfnamefont{O.}~\bibnamefont{Mesquita}},
  \bibinfo{author}{\bibfnamefont{J.}~\bibnamefont{Figueiredo}},
  \bibnamefont{and} \bibinfo{author}{\bibfnamefont{A.}~\bibnamefont{Vidal}},
  \bibinfo{journal}{Journal of Crystal Growth} \textbf{\bibinfo{volume}{166}},
  \bibinfo{pages}{222} (\bibinfo{year}{1996}).

\bibitem[{\citenamefont{Misbah and Valance}(1995)}]{Misbah1995}
\bibinfo{author}{\bibfnamefont{C.}~\bibnamefont{Misbah}} \bibnamefont{and}
  \bibinfo{author}{\bibfnamefont{A.}~\bibnamefont{Valance}},
  \bibinfo{journal}{Phys. Rev. E} \textbf{\bibinfo{volume}{51}},
  \bibinfo{pages}{1282} (\bibinfo{year}{1995}).

\bibitem[{\citenamefont{Gonz\'alez-Cinca
  et~al.}(1998)\citenamefont{Gonz\'alez-Cinca, Ram\'irez-Piscina, Casademunt,
  Hern\'andez-Machado, T\'oth-Katona, B\"orzs\"onyi, and
  Buka}}]{Gonzalez-Cinca1998}
\bibinfo{author}{\bibfnamefont{R.}~\bibnamefont{Gonz\'alez-Cinca}},
  \bibinfo{author}{\bibfnamefont{L.}~\bibnamefont{Ram\'irez-Piscina}},
  \bibinfo{author}{\bibfnamefont{J.}~\bibnamefont{Casademunt}},
  \bibinfo{author}{\bibfnamefont{A.}~\bibnamefont{Hern\'andez-Machado}},
  \bibinfo{author}{\bibfnamefont{T.}~\bibnamefont{T\'oth-Katona}},
  \bibinfo{author}{\bibfnamefont{T.}~\bibnamefont{B\"orzs\"onyi}},
  \bibnamefont{and} \bibinfo{author}{\bibfnamefont{A.}~\bibnamefont{Buka}},
  \bibinfo{journal}{Journal of Crystal Growth} \textbf{\bibinfo{volume}{193}},
  \bibinfo{pages}{712} (\bibinfo{year}{1998}),
  \urlprefix\url{http://www.sciencedirect.com/science/article/B6TJ6-3V7899C-2C%
/2/8bc63c20ec0948537e165a858a39b668}.

\bibitem[{\citenamefont{Bechhoefer}(1996)}]{Bechhoefer1996}
\bibinfo{author}{\bibfnamefont{J.}~\bibnamefont{Bechhoefer}},
  \emph{\bibinfo{title}{Pattern Formation in Liquid Crystals}}
  (\bibinfo{publisher}{Springer}, \bibinfo{year}{1996}), chap.
  \bibinfo{chapter}{Mesophase Growth}, pp. \bibinfo{pages}{257--283}.

\bibitem[{\citenamefont{Oswald et~al.}(1987)\citenamefont{Oswald, Bechhoefer,
  and Libchaber}}]{Oswald1987}
\bibinfo{author}{\bibfnamefont{P.}~\bibnamefont{Oswald}},
  \bibinfo{author}{\bibfnamefont{J.}~\bibnamefont{Bechhoefer}},
  \bibnamefont{and}
  \bibinfo{author}{\bibfnamefont{A.}~\bibnamefont{Libchaber}},
  \bibinfo{journal}{Phys. Rev. Lett.} \textbf{\bibinfo{volume}{58}},
  \bibinfo{pages}{2318} (\bibinfo{year}{1987}).

\bibitem[{\citenamefont{Flemmings}(1974)}]{Flemmings1974}
\bibinfo{author}{\bibfnamefont{M.}~\bibnamefont{Flemmings}},
  \emph{\bibinfo{title}{Solidification Processing}}
  (\bibinfo{publisher}{McGraw-Hill}, \bibinfo{year}{1974}).

\bibitem[{\citenamefont{Pimpinelli and Villain}(1998)}]{Pimpinelli1998}
\bibinfo{author}{\bibfnamefont{A.}~\bibnamefont{Pimpinelli}} \bibnamefont{and}
  \bibinfo{author}{\bibfnamefont{J.}~\bibnamefont{Villain}},
  \emph{\bibinfo{title}{Physics of Crystal Growth}}
  (\bibinfo{publisher}{Cambridge University Press}, \bibinfo{year}{1998}).

\bibitem[{\citenamefont{Narayan}(2002)}]{Narayan2002}
\bibinfo{author}{\bibfnamefont{J.}~\bibnamefont{Narayan}},
  \emph{\bibinfo{title}{Interfacial Stability}} (\bibinfo{publisher}{Springer},
  \bibinfo{year}{2002}).

\bibitem[{\citenamefont{Sethna}(2007)}]{Sethna2007}
\bibinfo{author}{\bibfnamefont{J.~P.} \bibnamefont{Sethna}},
  \emph{\bibinfo{title}{Statistical Mechanics: Entropy, Order Parameters and
  Complexity}} (\bibinfo{publisher}{Clarendon Press}, \bibinfo{year}{2007}),
  \urlprefix\url{http://www.us.oup.com/us/catalog/general/subject/Physics/Quan%
tumPhysics/?view=usa&ci=9780198566779}.

\bibitem[{\citenamefont{Xu}(1997)}]{Xu1997}
\bibinfo{author}{\bibfnamefont{J.-J.} \bibnamefont{Xu}},
  \emph{\bibinfo{title}{Interfacial Wave Theory of Pattern Formation: Selection
  of Dendritic Growth and Viscous Fingering in Hele-Shaw Flow}}
  (\bibinfo{publisher}{Springer}, \bibinfo{year}{1997}).

\bibitem[{\citenamefont{Pismen}(2006)}]{Pismen2006}
\bibinfo{author}{\bibfnamefont{L.~M.} \bibnamefont{Pismen}},
  \emph{\bibinfo{title}{Patterns and Interfaces in Dissipative Dynamics}}
  (\bibinfo{publisher}{Springer}, \bibinfo{year}{2006}).

\bibitem[{\citenamefont{de~Gennes and Prost}(1995)}]{deGennes1995}
\bibinfo{author}{\bibfnamefont{P.}~\bibnamefont{de~Gennes}} \bibnamefont{and}
  \bibinfo{author}{\bibfnamefont{J.}~\bibnamefont{Prost}},
  \emph{\bibinfo{title}{The Physics of Liquid Crystals}}
  (\bibinfo{publisher}{Oxford University Press}, \bibinfo{address}{New York},
  \bibinfo{year}{1995}), \bibinfo{edition}{2nd} ed.

\bibitem[{\citenamefont{Gramsbergen et~al.}(1986)\citenamefont{Gramsbergen,
  Longa, and de~Jeu}}]{Gramsbergen1986}
\bibinfo{author}{\bibfnamefont{E.~F.} \bibnamefont{Gramsbergen}},
  \bibinfo{author}{\bibfnamefont{L.}~\bibnamefont{Longa}}, \bibnamefont{and}
  \bibinfo{author}{\bibfnamefont{W.~H.} \bibnamefont{de~Jeu}},
  \bibinfo{journal}{Physics Reports} \textbf{\bibinfo{volume}{135}},
  \bibinfo{pages}{195} (\bibinfo{year}{1986}),
  \urlprefix\url{http://www.sciencedirect.com/science/article/B6TVP-46P3WM2-3T%
/1/46e0a6a1464a3f70d653423446277512}.

\bibitem[{\citenamefont{Singh}(2000)}]{Singh2000}
\bibinfo{author}{\bibfnamefont{S.}~\bibnamefont{Singh}},
  \bibinfo{journal}{Physics Reports} \textbf{\bibinfo{volume}{324}},
  \bibinfo{pages}{107} (\bibinfo{year}{2000}),
  \urlprefix\url{http://www.ingentaconnect.com/content/els/03701573/2000/00000%
324/00000002/art00049}.

\bibitem[{\citenamefont{Kibble}(2007)}]{Kibble2007}
\bibinfo{author}{\bibfnamefont{T.}~\bibnamefont{Kibble}},
  \bibinfo{journal}{Physics Today} \textbf{\bibinfo{volume}{60}},
  \bibinfo{pages}{47} (\bibinfo{year}{2007}).

\bibitem[{\citenamefont{Rey}(2007)}]{Rey2007}
\bibinfo{author}{\bibfnamefont{A.~D.} \bibnamefont{Rey}},
  \bibinfo{journal}{Soft Matter} \textbf{\bibinfo{volume}{3}},
  \bibinfo{pages}{1349 } (\bibinfo{year}{2007}).

\bibitem[{\citenamefont{Abukhdeir et~al.}(2008)\citenamefont{Abukhdeir,
  Soul\'{e}, and Rey}}]{Abukhdeir2008b}
\bibinfo{author}{\bibfnamefont{N.~M.} \bibnamefont{Abukhdeir}},
  \bibinfo{author}{\bibfnamefont{E.~R.} \bibnamefont{Soul\'{e}}},
  \bibnamefont{and} \bibinfo{author}{\bibfnamefont{A.~D.} \bibnamefont{Rey}},
  \bibinfo{journal}{Langmuir}  (\bibinfo{year}{2008}), ISSN
  \bibinfo{issn}{0743-7463}, \bibinfo{note}{web Release Date: October 29, 2008;
  DOI: 10.1021/la8022216},
  \urlprefix\url{http://pubs3.acs.org/acs/journals/doilookup?in_doi=10.1021/la%
8022216}.

\bibitem[{\citenamefont{Oswald and Pieranski}(2005)}]{Oswald2005}
\bibinfo{author}{\bibfnamefont{P.}~\bibnamefont{Oswald}} \bibnamefont{and}
  \bibinfo{author}{\bibfnamefont{P.}~\bibnamefont{Pieranski}},
  \emph{\bibinfo{title}{Nematic and cholesteric liquid crystals : concepts and
  physical properties illustrated by experiments}}, Liquid crystals book series
  (\bibinfo{publisher}{Taylor \& Francis/CRC Press}, \bibinfo{address}{Boca
  Raton, FL}, \bibinfo{year}{2005}).

\bibitem[{\citenamefont{Pestov}(2003)}]{Pestov2003}
\bibinfo{author}{\bibfnamefont{S.}~\bibnamefont{Pestov}},
  \emph{\bibinfo{title}{Landolt-Bernstein - Group VIII Advanced Materials and
  Technologies}} (\bibinfo{publisher}{Springer Berlin / Heidelberg},
  \bibinfo{year}{2003}), chap. \bibinfo{chapter}{2.1.1 Two ring systems without
  bridge},
  \urlprefix\url{http://www.springerlink.com/content/w92mtf6n7v6fn8av/}.

\bibitem[{\citenamefont{Rey and Denn}(2002)}]{Rey2002}
\bibinfo{author}{\bibfnamefont{A.}~\bibnamefont{Rey}} \bibnamefont{and}
  \bibinfo{author}{\bibfnamefont{M.}~\bibnamefont{Denn}},
  \bibinfo{journal}{Annual Review of Fluid Mechanics}
  \textbf{\bibinfo{volume}{34}}, \bibinfo{pages}{p233 } (\bibinfo{year}{2002}),
  ISSN \bibinfo{issn}{00664189},
  \urlprefix\url{http://search.ebscohost.com/login.aspx?direct=true&db=aph&AN=%
6262786&site=ehost-live}.

\bibitem[{\citenamefont{Yan and Rey}(2002)}]{Yan2002}
\bibinfo{author}{\bibfnamefont{J.}~\bibnamefont{Yan}} \bibnamefont{and}
  \bibinfo{author}{\bibfnamefont{A.~D.} \bibnamefont{Rey}},
  \bibinfo{journal}{Phys. Rev. E} \textbf{\bibinfo{volume}{65}},
  \bibinfo{pages}{031713} (\bibinfo{year}{2002}).

\bibitem[{\citenamefont{Wincure and Rey}(2007{\natexlab{b}})}]{Wincure2007}
\bibinfo{author}{\bibfnamefont{B.}~\bibnamefont{Wincure}} \bibnamefont{and}
  \bibinfo{author}{\bibfnamefont{A.}~\bibnamefont{Rey}},
  \bibinfo{journal}{Continuum Mechanics and Thermodynamics}
  \textbf{\bibinfo{volume}{19}}, \bibinfo{pages}{37}
  (\bibinfo{year}{2007}{\natexlab{b}}).

\bibitem[{\citenamefont{Wincure and Rey}(2007{\natexlab{c}})}]{Wincure2007b}
\bibinfo{author}{\bibfnamefont{B.}~\bibnamefont{Wincure}} \bibnamefont{and}
  \bibinfo{author}{\bibfnamefont{A.~D.} \bibnamefont{Rey}},
  \bibinfo{journal}{Liquid Crystals} \textbf{\bibinfo{volume}{34}},
  \bibinfo{pages}{1397} (\bibinfo{year}{2007}{\natexlab{c}}), ISSN
  \bibinfo{issn}{0267-8292},
  \urlprefix\url{http://www.informaworld.com/10.1080/02678290701614657}.

\bibitem[{\citenamefont{Barbero and Evangelista}(2000)}]{Barbero2000}
\bibinfo{author}{\bibfnamefont{G.}~\bibnamefont{Barbero}} \bibnamefont{and}
  \bibinfo{author}{\bibfnamefont{L.~R.} \bibnamefont{Evangelista}},
  \emph{\bibinfo{title}{An Elementary Course on the Continuum Theory for
  Nematic Liquid Crystals (Series on Liquid Crystals , Vol 3)}}
  (\bibinfo{publisher}{World Scientific Publishing Company},
  \bibinfo{year}{2000}), ISBN \bibinfo{isbn}{9810232241}.

\bibitem[{\citenamefont{Kaiser et~al.}(1992)\citenamefont{Kaiser, Wiese, and
  Hess}}]{Kaiser1992}
\bibinfo{author}{\bibfnamefont{P.}~\bibnamefont{Kaiser}},
  \bibinfo{author}{\bibfnamefont{W.}~\bibnamefont{Wiese}}, \bibnamefont{and}
  \bibinfo{author}{\bibfnamefont{S.}~\bibnamefont{Hess}}, \bibinfo{journal}{J.
  Non-Equilib. Thermodyn} \textbf{\bibinfo{volume}{17}}, \bibinfo{pages}{153}
  (\bibinfo{year}{1992}).

\bibitem[{\citenamefont{Kralj et~al.}(1999)\citenamefont{Kralj, Virga, and
  Zumer}}]{Kralj1999}
\bibinfo{author}{\bibfnamefont{S.}~\bibnamefont{Kralj}},
  \bibinfo{author}{\bibfnamefont{E.~G.} \bibnamefont{Virga}}, \bibnamefont{and}
  \bibinfo{author}{\bibfnamefont{S.}~\bibnamefont{Zumer}},
  \bibinfo{journal}{Phys. Rev. E} \textbf{\bibinfo{volume}{60}},
  \bibinfo{pages}{1858} (\bibinfo{year}{1999}).

\bibitem[{\citenamefont{Yokoyama}(1997)}]{Yokoyama1997}
\bibinfo{author}{\bibfnamefont{H.}~\bibnamefont{Yokoyama}},
  \emph{\bibinfo{title}{Handbook of Liquid Crystal Research}}
  (\bibinfo{publisher}{Oxford University Press}, \bibinfo{year}{1997}), chap.
  \bibinfo{chapter}{Interfaces and Thin Films}, p. \bibinfo{pages}{179}.

\end{thebibliography}

\end{document}